%% file: main.tex
\documentclass[twocolumn]{aastex631}
\usepackage{amsmath}
\usepackage{booktabs}
\usepackage{xspace}
\usepackage{hyperref}

\newcommand{\halpha}{H$\alpha$}
\newcommand{\hbeta}{H$\beta$}
\newcommand{\hgamma}{H$\gamma$}

\newcommand{\mdot}{$\dot{\text{M}}$}
\newcommand{\Mdot}{{\dot{{M}}}}

\newcommand{\rsun}{ R_{\sun}}

\newcommand{\msunyr}{M_{\sun} \, \rm{ yr^{-1}}}

\newcommand{\teff}{T$_{\rm eff}$}

\newcommand{\ri}{R$_{\rm i}$}

\newcommand{\tmax}{T$_{\rm max}$}

\begin{document}

\title{Understanding Balmer Decrements in T Tauri stars in terms of Multiflow Magnetospheric Accretion}

\correspondingauthor{Naiara Patiño}
\email{nzpatino@bu.edu}

\author[0009-0009-7455-6777]{Naiara Patiño}
\affiliation{Institute for Astrophysical Research, Department of Astronomy, Boston University, 725 Commonwealth Avenue, Boston, MA 02215, USA}

\author[0000-0002-3950-5386]{Nuria Calvet}
\affiliation{Department of Astronomy, University of Michigan, 1085 South University Avenue, Ann Arbor, MI 48109, USA}

\author[0000-0003-1166-5123]{Gladis Magris}
\affiliation{Centro de Investigaciones de Astronomía (CIDA), Mérida, 5101, Venezuela.}

\author[0000-0001-8022-4378]{Marbely Micolta}
\affiliation{Department of Astronomy, University of Michigan, 1085 South University Avenue, Ann Arbor, MI 48109, USA}

\author[0000-0003-4507-1710]{Thanawuth Thanathibodee}
\affiliation{Department of Physics, Faculty of Science, Chulalongkorn University, 254 Phayathai Road, Pathumwan, Bangkok, 10330 Thailand.}

\author[0000-0002-5231-7240]{Thomas K. Waters}
\affiliation{Department of Astronomy, University of Michigan, 1085 South University Avenue, Ann Arbor, MI 48109, USA}

\author[0000-0002-5296-6232]{María José Colmenares}
\affiliation{Department of Astronomy, University of Michigan, 1085 South University Avenue, Ann Arbor, MI 48109, USA}

\shorttitle{Understanding Balmer decrements in CTTS using multi-flow magnetospheric accretion}
\shortauthors{Patiño et al.}

\begin{abstract}

Magnetospheric accretion is the paradigm for accretion in Classical T-Tauri Stars (CTTS). However, the standard, one-flow magnetospheric accretion model fails to replicate important characteristics such as the observed Balmer decrements. We address this limitation by adopting a model with two axisymmetric magnetospheric accretion flows of different accretion rates and geometries. We calculate the fluxes of the hydrogen {\halpha}, {\hbeta}, and \hgamma\ lines of each flow with the magnetospheric accretion model and use Bayesian statistics to fit the Balmer line fluxes of 139 CTTS in the Orion OB1b subassociation, and in the Upper Scorpius, Lupus and Chamaeleon I star-forming regions. We find that the Balmer decrements and line fluxes can be fitted by two distinct but coexisting flows: a compact, high accretion rate flow, close to the star and narrow (mean inner radius $R_i \sim 2.9 R_*$ and mean width $\Delta R \sim 0.7 R_*$), covering a few percent of the emitting area, and a more spread out flow, thicker ($\Delta R \sim 1.2 R_*$), and larger ($R_i \sim 3.7 R_*$), with lower accretion rate, encompassing the rest of the emitting area. The two-flow model can also reproduce the empirical correlation between the luminosity in \halpha\ and the accretion luminosity. Overall, our findings suggest that a multicolumn approach provides a more accurate representation of the observed Balmer line emission, in agreement with results of numerical simulations.

\end{abstract}


\section{Introduction}

Conservation of angular momentum leads to the development of a disk of gas and dust during the formation of a star from the collapse of a molecular cloud core. Young stars accrete mass from this disk, and this process plays a crucial role in shaping the early evolution of both star and disk \citep[see][for a review]{hartmann2016}.

Some of the best studied young stellar objects are the T Tauri stars (TTS), which are low-mass stars of late spectral types \citep{hartmann2016}. These objects can be further classified into accreting Classical T Tauri Stars (CTTS) and non-accreting Weak-lined T Tauri Stars (WTTS). Investigating the physical processes at play in these systems is essential for understanding the evolution of protoplanetary disks and the formation of planets. The prevailing model for mass accretion in TTS is the magnetospheric accretion paradigm \citep{hartmann2016, muzerolle1998a, muzerolle1998b, muzerolle2001, hartmann1994}.  In this framework, the stellar magnetic field truncates the inner disk, channeling disk material along field lines in accretion columns onto the stellar surface. Several lines of evidence support this model, including the observed ultraviolet (UV) excess produced by accretion shocks at the stellar surface, where infalling material decelerates and merges with the stellar atmosphere \citep{calvet_gullbring1998}, as well as the broad wings observed in emission lines \citep{muzerolle2001}. The excess continuum, and the profile width and flux of emission lines are widely used to estimate mass accretion rates in TTS \citep{hartmann1994, muzerolle1998a, muzerolle1998b, muzerolle2001}.

Despite its success, \citet{muzerolle2001} found that the magnetospheric accretion model tends to underpredict
the Balmer lines ratio {\hgamma /\hbeta}.

They attributed this discrepancy to an inherent limitation of the model. One of the core assumptions of the magnetospheric accretion model is that of axisymmetry and homogeneity of the stellar magnetic field. However, both observational studies \citep[e.g.,][]{espaillat2021} and magnetohydrodynamic (MHD) simulations \citep[e.g.,][]{zhaohuan2024, romanova2003} indicate that the accretion flows are typically neither axisymmetric nor homogeneous. Here, we explore the implications of considering an inhomogeneous magnetosphere.

In this work, we approximate the complex magnetospheric geometry expected in CTTS with a model consisting of two distinct accretion flows with different parameters, aiming to
explain the observed hydrogen emission lines. The concept of multicolumn accretion has previously been applied to accretion shock models; columns of different energy density - or volume mass density, since the model adopts free-fall velocities - can
successfully account for both near-ultraviolet (NUV) and optical excess emission in CTTS \citep{ingleby2013, pittman2022}. There is also evidence for density gradients in the hot spots produced by accretion columns on the stellar surface \citep{espaillat2021}, agreeing with the interpretation of multiple accretion streams of varying density. Here, we extend this concept by studying multicolumn accretion, focusing on the line emission originating from the extended accretion flows themselves rather than from the shock on the stellar surface. 

We analyze CTTS in four star-forming regions: Lupus \citep{alcala2014, alcala2017},  Chamaeleon I \citep{manara2016, manara2017b}, Orion OB1b \citep{briceno2019}, and Upper Scorpius \citep{manara2020} (hereafter, Lup, Cha I, OB1b, and USco, respectively). These regions offer a diverse sample in terms of age, stellar parameters, and evolutionary stages, enabling us to probe accretion across different environments.

This paper is organized as follows. In Section \ref{sec:Sample and observations}, we describe the sample of T Tauri stars, their properties, and the observational data used. Section \ref{Models} presents the standard single-column magnetospheric accretion model and its limitations, introduces our proposed multicolumn model, and outlines the spectral analysis, correction procedures, and statistical methods employed. Section \ref{decrements} reports our findings on the impact of this modeling approach on Balmer decrements, while Section \ref{individual_stars} details the results obtained 
from modeling the line fluxes of the stars in the samples.
Finally, we discuss the broader
implications of our results in Section \ref{Discussion}.

\section{Sample} 
\label{sec:Sample and observations}

For all of our targets, we use spectra obtained with the X-Shooter spectrograph on the Very Large Telescope (VLT) \citep{vernet2011}. The X-Shooter spectrograph offers medium resolution,
broad wavelength coverage, and flux calibration. Its spectral range spans $\sim$300–2500nm, enabling the extraction of key stellar parameters, as well as diagnostics of the accretion process and inner disk structure. 

\subsection{Classical T Tauri Stars} 

\begin{figure*}
    \centering
    \includegraphics[width=\linewidth]{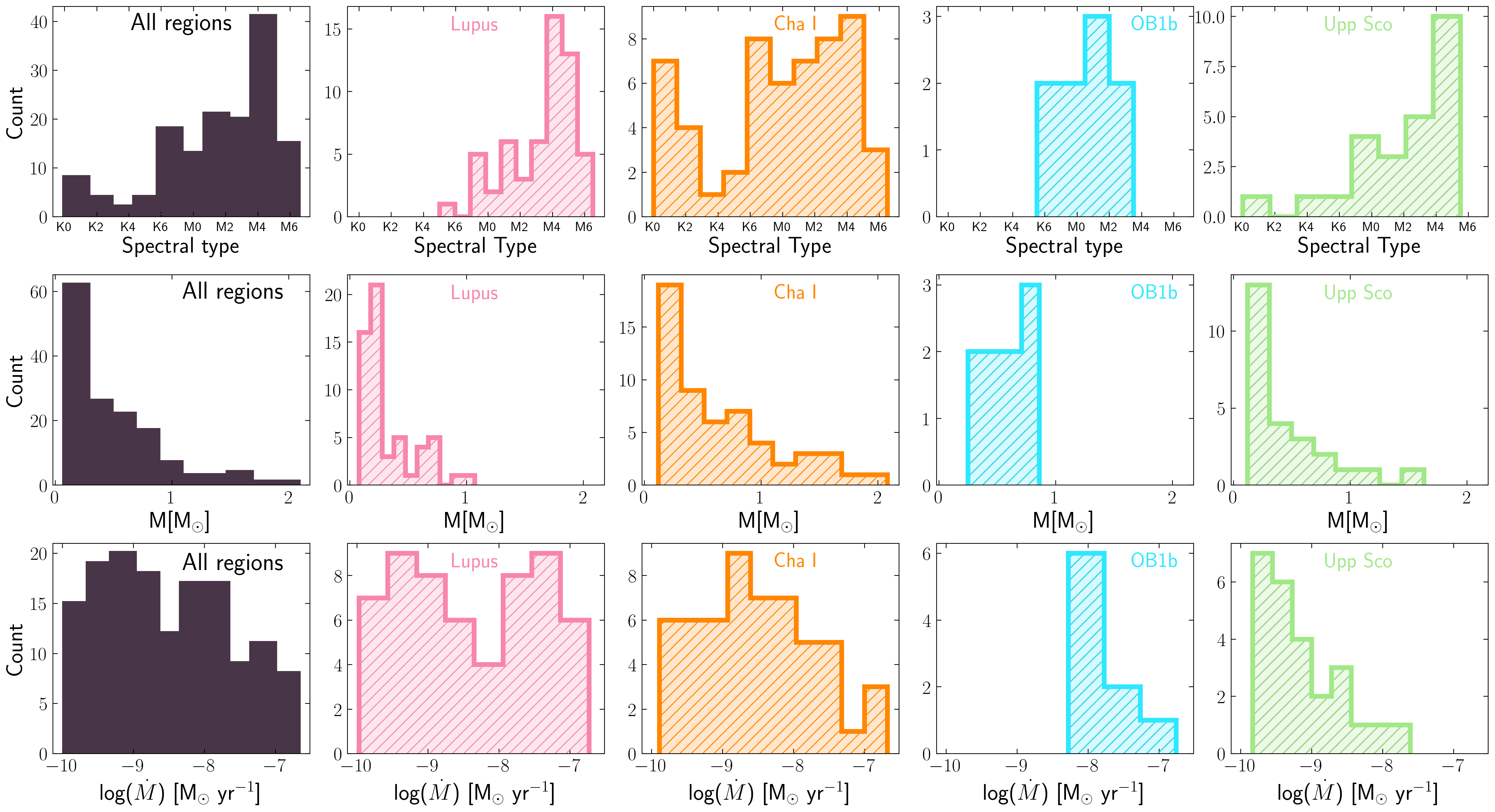}
    \caption{From left to right, distribution of relevant stellar parameters of the total sample, Lup, Cha I, Ori OB1b, and USco. These parameters are as reported in \citet{alcala2014,alcala2017} (Lup), \citet{manara2016, manara2017b} (Cha I),  \citet{pittman2022} for \mdot\ and \citet{manara2021} for the rest (OB1b), and \citet{manara2020} (USco). Note that stellar ages increase from left to right. \textit{Top:} Distribution of spectral types in our sample. \textit{Middle:} Distribution of stellar masses.  \textit{Bottom:} Distribution of accretion rates.}
    \label{fig:params_CTTS}
\end{figure*}

The key stellar and accretion parameters adopted for the final sample of 139 CTTS are shown in \autoref{fig:params_CTTS}.  
We took the parameters from \citet{alcala2014,alcala2017}(Lup), \citet{manara2016, manara2017b} (Cha I),  \citet{pittman2022} and \citet{manara2021} (OB1b), and \citet{manara2020} (USco). A complete table listing all parameters extracted from the literature is provided in \autoref{table:ctts_table}.

Our sample focused on targets with accretion rates lower than $10^{-7}\msunyr$ and that did not exhibit a “P Cygni” profile in their Balmer lines, characteristic of strong winds and/or jets \citep[e.g. see Figure 11 of][]{muzerolle2001}. This way any contributions to the flux due to winds, which our model does not account for, can be rule out.

Fluxes for the \halpha, \hbeta, and \hgamma\ lines in Lupus and Cha I were taken directly from the literature. For Upper Scorpius and Orion OB1b we calculated the line fluxes by integrating the dereddened, heliocentric-velocity-corrected spectra after subtracting the continuum, which was determined using the \texttt{fit\_generic\_continuum} function from the \texttt{specutils} Python package. Flux uncertainties were estimated following the method described by \citet{alcala2014}. This approach involves defining two additional continua—one above and one below the original—and computing the line flux for each. The final flux is taken as the average of the three values, and the uncertainty is given by their standard deviation, $\sigma$. In our case, the alternative continua were defined by varying the original continuum by $\pm$ 10 \%. 

\smallskip

In the following sections, we describe the most relevant and distinctive characteristics of each region, in order of increasing estimated age. 

\bigskip

\subsubsection{Lupus}

The Lupus region is one of the nearest star-forming regions ($d < 200$ pc) and hosts objects at various evolutionary stages \citep[see][for a review]{comeron2008}. Our Lupus sample 
of 92 CTTS, includes 36 objects from \citet{alcala2014} and 55 from \citet{alcala2017}. Stellar and accretion parameters, extinction coefficients, and line fluxes, were taken directly from \citet{alcala2014, alcala2017}. In these works, accretion parameters were derived using a slab model: the observed spectra were fitted using a combination of a WTTS photospheric template and a slab model that simulates accretion shock emission \citep[see][]{valenti1993, HerczegHillenbrand2008,alcala2014, alcala2017,manara2021,manara2020}. 

\citet{alcala2014, alcala2017} estimated the average age of the Lupus sample to be 3 $\pm$ 2 Myr. From the original sample, we excluded non-accreting or low-accreting sources ($\dot{M} < 10^{-10} M_\odot\,\mathrm{yr}^{-1}$), those not confirmed to be members of the Lupus region, as well as those lacking measured \halpha, \hbeta, or \hgamma\ fluxes. This selection resulted in a final sample of 56 objects.

\subsubsection{Chamaeleon I}
The Chamaeleon I sample consists of 51 Class II objects studied in \citet{manara2016, manara2017b}, from which we took stellar and accretion parameters, extinction estimates, and line fluxes. 
Mass accretion rates were obtained in these works using a slab model. The sources included from this region exhibit low visual extinctions ($A_V \lesssim 3$ mag). In this work, we adopt a distance of $d = 160$ pc for Chamaeleon I, and an age estimate of 2 Myr from \citet{luhman2008}.

\subsubsection{Orion OB1b} \label{ob1b}

The Orion OB1 association is one of the nearest and most populous OB associations, with subassociations spanning ages from $\sim$1 to 10 Myr \citep{hernandez2007, briceno2019}. 
We analyzed a subsample of nine CTTS in the Orion OB1b subassociation from the study of \citet{manara2021}, after excluding systems with accretion rates below $10^{-10} M_\odot \mathrm{yr}^{-1}$. We adopted spectral types, stellar masses, and distances reported therein.
We adopted stellar radii, mass accretion rates, and V-band extinctions from \citet{pittman2022}, derived using a Markov Chain Monte Carlo (MCMC) fitting procedure based on a three-column magnetospheric accretion shock model. The estimated age of the subassociation is $\sim$ 5 Myr \citep{briceno2019} 

\subsubsection{Upper Scorpius}

We selected our Upper Scorpius sample from the 36 young stellar objects analyzed by \citet{manara2020}, from which we also adopted the stellar and accretion parameters. From this sample, we excluded sources identified as non-accreting or with low accretion rates ($\Mdot < 10^{-10} $M$_\odot \mathrm{yr}^{-1}$) based on slab modeling by \citet{manara2020}. This selection resulted in a final sample of 23 CTTS in Upper Scorpius.

Upper Scorpius is the oldest region in our study, with age estimates of $\sim$5–10 Myr \citep[e.g.,][]{pecaut2012, preibisch2002, sullivan_kraus2021}, and it exhibits the lowest accretion rates among the regions considered \citep{manara2020}.

\subsection{Weak T Tauri Stars}\label{chromospheric_emission}

Due to the high surface magnetic activity, the stellar chromosphere emits significantly in the UV and in emission lines even in non-accreting stars. We used WTTS as templates to account for non-accretion-related flux excess due to chromospheric emission in CTTS. These templates were taken from \citet{manara2013b, manara2017a}, which include non-accreting objects from nearby star-forming regions ($d < 500$ pc) with negligible visual extinctions ($A_V < 0.5$ mag), and measurements in multiple photometric bands. The emission fluxes for \halpha\, \hbeta\, and \hgamma\, were obtained from \citet{micolta2023}. The stellar parameters of the final WTTS sample are given in \autoref{table:wtts_table}.

\section{Model} \label{Models}

\begin{deluxetable}{lllll}
    \tablecaption{Model stellar parameters} 
    \label{table:parametros_modelos}
    \tablehead{\colhead{SpT} & \colhead{Teff [K]} & \colhead{L [$L_{\odot}$]} & \colhead{R [$R_{\odot}$]} & \colhead{M [$M_{\odot}$]}}
\startdata
    K5  & 4350         & 0.745               & 1.519               & 0.872               \\
    K7  & 4060         & 0.515               & 1.451               & 0.726               \\
    M1  & 3705         & 0.345               & 1.426               & 0.614               \\
    M3  & 3415         & 0.215               & 1.327               & 0.465               \\
    M5  & 3125         & 0.117               & 1.171               & 0.306               \\
\enddata
    \tablecomments{Stellar parameters for the models considered. These values include spectral type (SpT),  effective temperature (\teff), luminosity ($L$), radius ($R$), and mass ($M$) of the stars.}
\end{deluxetable}

We use the magnetospheric accretion model \citep{bertout1988,camenzind1990, konigl1991}  as implemented by \citet{hartmann1994}
and \citet{muzerolle1998a, muzerolle1998b, muzerolle2001} to characterize the emission of the accretion flows. This model assumes an axisymmetric and dipolar stellar magnetic field, with the stellar rotation axis aligned with its magnetic pole. 
The disk is truncated by the magnetic field at a radius R$_i$ and material from the disk is channeled onto the star by the field lines. 
In our models, the geometry of these flows is defined by an internal radius (R$_i$), which corresponds to the disk's truncation radius, and an external radius (R$_o$), which in turn determines the width of the flows at their base ($\Delta$R). Each model assumes a constant mass accretion rate (\mdot), from which the density structure is calculated. The temperature is calculated equating a parametric heating law to optically thin cooling \citep{hartmann1994} and each model is characterized by the maximum temperature in the accretion flows (\tmax). We calculate level populations for a 20-level H atom. Mean intensities for each radiative transition are calculated with the extended Sobolev approximation following \citet{hartmann1994} and \citet{muzerolle2001}. Line fluxes are obtained using a ray-by-ray integration, with optical depths and source functions calculated from the H populations, and using Voigt profiles \citep{muzerolle2001}.

\begin{deluxetable}{llll}
\tablecaption{Parameter space for accretion models}
\tablehead{
\colhead{Parameter} & \colhead{Min.} & \colhead{Max.} & \colhead{Step}
}
\label{table:parameter_space}
\startdata
log($\Mdot$) [$\msunyr$] & -10  & -6.5    & 0.25 \\
$T_{max}$ [$K$]          & 6500 & 14000 & 500  \\
\ri [$\rsun$]          & 2    & 6     & 0.5  \\
$\Delta$R [$\rsun$]          & 0.5  & 2     & 0.5  \\
$i$ [$^\circ$]           & 15   & 75    & 15   \\
\enddata
\tablecomments{The table above lists the range and step of parameters used in the accretion model simulations. The parameters are the log mass accretion rate (log($\Mdot$)), the maximum temperature ($T_{max}$), the inner disk radius (\ri), the thickness of the disk ($\Delta$R), and the inclination angle of the dipolar magnetic field with respect to the line of sight ($i$).}
\end{deluxetable}

We calculate profiles and fluxes for the H Balmer lines for a model grid of 5 stars with stellar parameters shown in Table \ref{table:parametros_modelos}, which cover characteristic values for CTTS in the sample (see \autoref{fig:params_CTTS}). For each model star, we obtained a grid of magnetospheric accretion models with parameters in Table \ref{table:parametros_modelos}, resulting in 91194 models. This grid has been used by \citet{micolta2023,micolta2024} to study the Ca abundance in the magnetospheric flows of CTTS. Here, we concentrate on the hydrogen lines \halpha, \hbeta, and \hgamma. A detailed analysis of upper-level Balmer lines is beyond the scope of the present work since it would require a substantial extension of both the modeling and the observational analysis. In addition, higher-order Balmer lines are not detected in all objects in the sample.

\begin{figure*}
    \centering
    \includegraphics[width=0.9\linewidth]{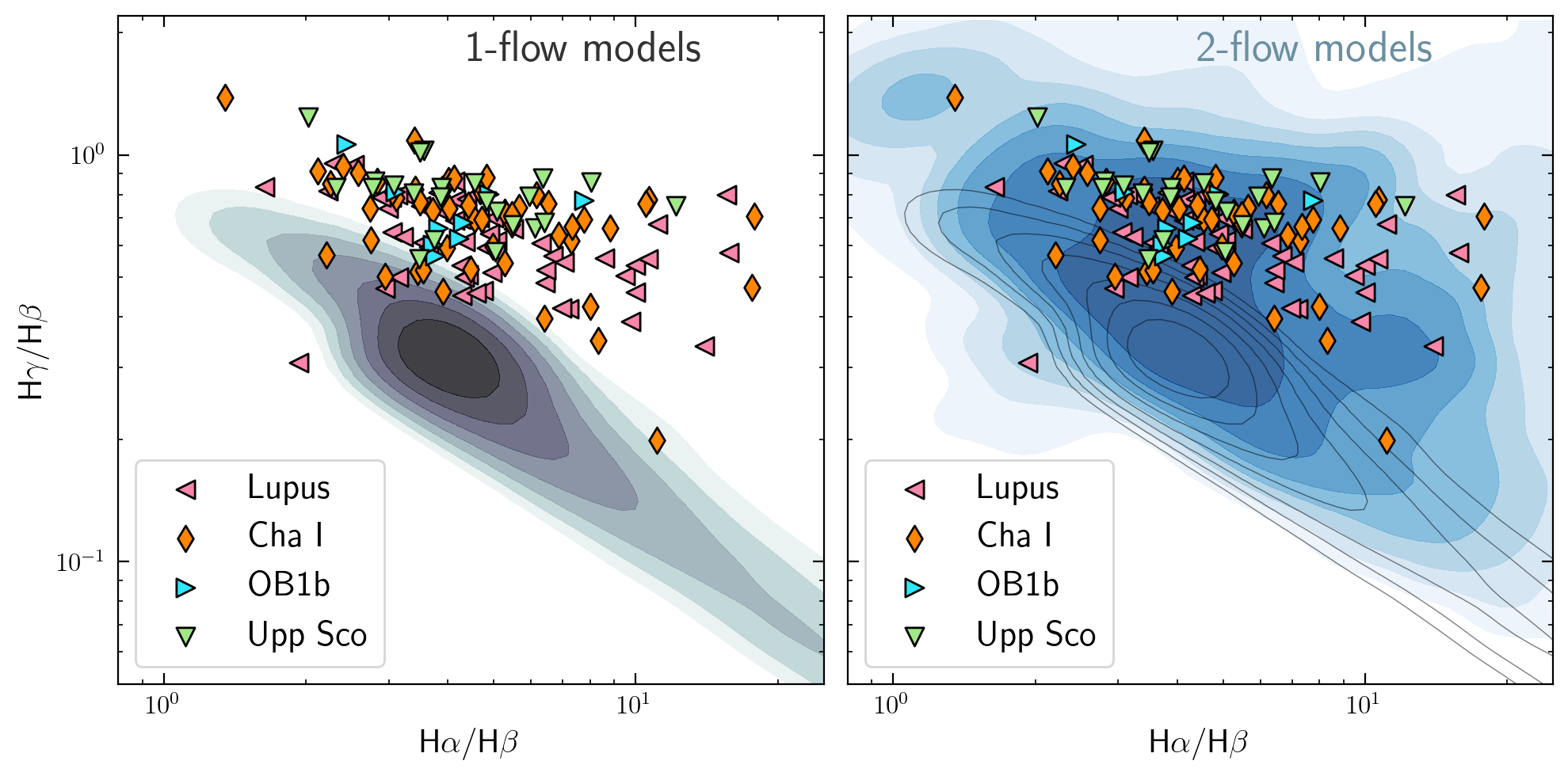}
    \caption{
    Predicted Balmer decrements by magnetospheric accretion models compared to observations. 
    The filled contours show the density of model predictions, computed in 100 logarithmically spaced bins, with darker shading indicating a higher concentration of models per bin. To improve readability, the model distributions have been smoothed using a Gaussian filter. 
    The scattered markers in both panels represent the Balmer decrements of CTTS in Lupus (pink), Chamaeleon I (orange), Orion OB1b (cyan), and Upper Scorpius (green), after subtracting the estimated chromospheric contribution. 
    \textit{Left panel}: Density contours for the Balmer decrements predicted by the single-column models. \textit{Right panel:} Same as the left, but using fluxes from the best-fit two-column models (details in Section \ref{individual_stars}). For comparison, the density contours from the single-column models (left panel) are overplotted as continuous grey lines.}
    \label{fig:balmer_decrement}
\end{figure*}

The left panel of \autoref{fig:balmer_decrement} shows the ratios \hgamma/\hbeta\ and \halpha/\hbeta\ predicted by models covering parameters in \autoref{table:parameter_space} for the model stars in Table \ref{table:parametros_modelos}.
\autoref{fig:balmer_decrement} also shows the redenning-corrected Balmer ratios of our observational sample. The model predictions do not agree with the observed Balmer decrements, indicating that the current model has intrinsic limitations. This was first shown by \citet{muzerolle2001} 
comparing model predictions with observations for CTTS from the Taurus star-forming region. In the following sections, we examine and relax some of the assumptions of the model, aiming to get a better understanding of the magnetospheric accretion phenomenon.

\subsection{Multi-flow models} \label{multicolumn_accretion}

We propose using a multi-flow model to study the origin of the discrepancies between observations and previously used accretion models. Non-uniform magnetospheric accretion has been explored in various contexts. For example, \citet{ingleby2013, espaillat2021,pittman2022,pittman2025} successfully explained excess emission over the photosphere by invoking accretion shocks produced by multiple columns with different energy densities. Similarly, \citet{atom2019b} used two accretion flows arranged in an onion-like structure to reproduce the observed \halpha\ profiles of a low accretor.

Here, as a proof of concept, we consider a model consisting of two magnetospheric flows with mass accretion rates $\Mdot_1$ and $\Mdot_2$, distinct maximum temperatures and geometries and emitting area, but the same inclination. We test whether the observed Balmer decrements can be reproduced by combining the line fluxes from these two columns. 
We assume that each column spans a fraction of the total emitting area described by the coverage factor $f_i$. The total flux $F_{l,\rm{comb}}$ predicted by our model for each line $l$ is defined in \autoref{eq:fmodcombined} as the sum of the fluxes from both accretion columns, each weighted by its corresponding coverage factor

\begin{equation}
    F_{l,\rm{comb}} = \frac{f_1  F_1 + f_2   F_2}{f_1 + f_2}.
    \label{eq:fmodcombined}
\end{equation}
Similarly, the total \mdot\ of each composite model is the sum of mass accretion rates of the two columns, weighted by their respective coverage factors,

\begin{equation}
    \Mdot_{\rm{comb}} = \frac{f_1 \Mdot_1 + f_2 \Mdot_2}{f_1 + f_2},
\end{equation}
where $\Mdot_{\rm{comb}}$ is the combined mass accretion rate of the two columns and $\Mdot_{i}$ is the mass accretion rate of the column corresponding to model $i$ and factor $f_i$. 

In our setup, we fix $f_1 = 10^{-3}$ and explore different values for the second column: $f_2 = 10^{-1}, 10^{-2}, 10^{-3}, 10^{-4}$. These values were chosen based on the typical filling factors used in accretion shock models \citep{calvet_gullbring1998}, which quantify the area of the stellar surface covered by the accretion shock emission. However, we note that the filling factors in accretion shock models and the $f$ values in our multi-flow accretion model used here are not equivalent, though both represent emitting area coverage. 
\bigskip

\begin{figure*}
    \centering
    \includegraphics[width=\linewidth]{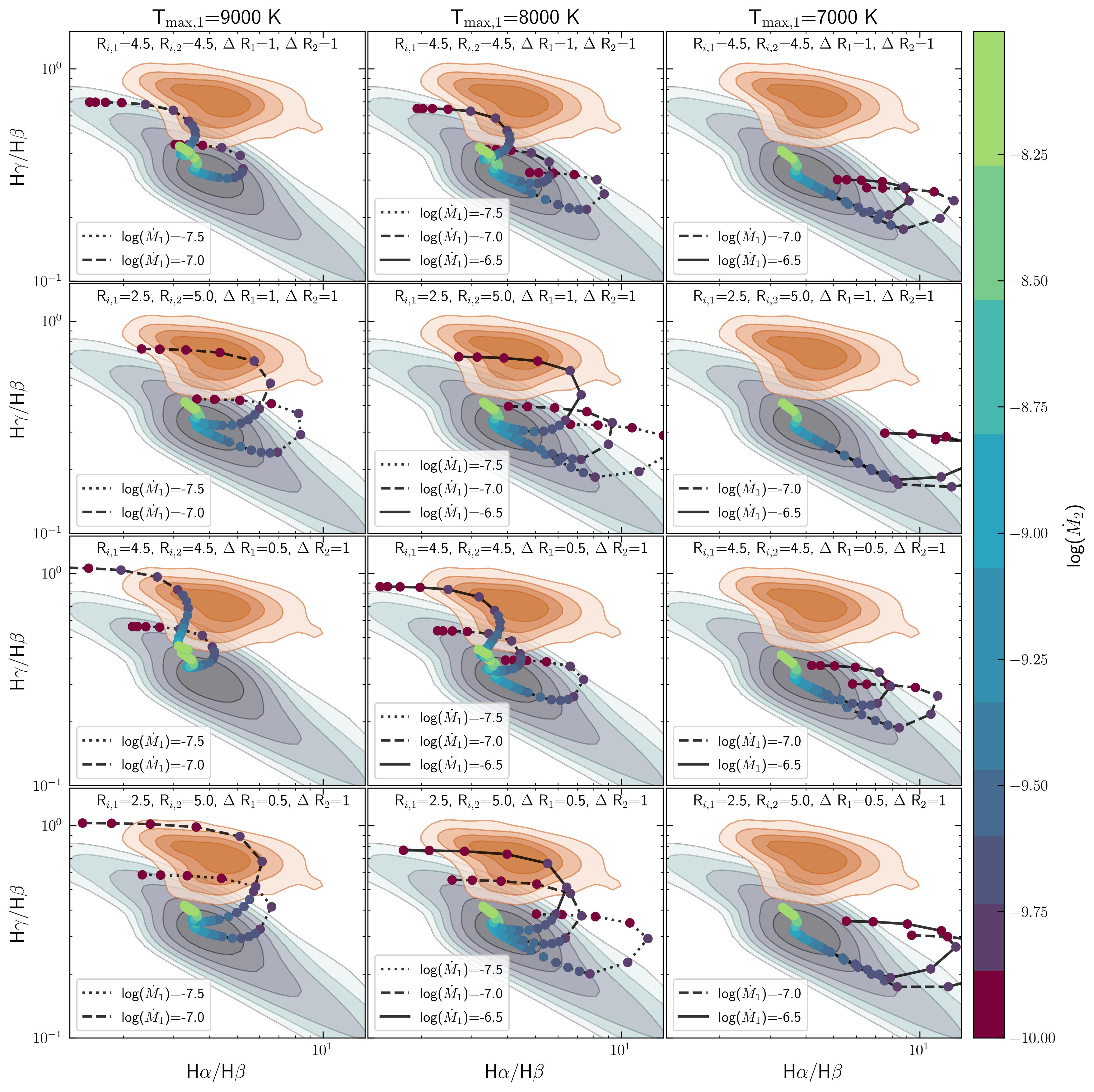}
    \caption{Examples of the effects of two accretion flows on Balmer decrements across different model parameters. From left to right, columns correspond to fixed maximum temperatures in the low-coverage column: $T_{\mathrm{max},1} = 10000$K, 9000K, and 7500K, respectively. Subscripts 1 and 2 refer to the low- and high-coverage columns, respectively, with fixed factors of $f_1 = 10^{-3}$ and $f_2 = 10^{-1.5}$. Each row illustrates a different magnetospheric geometry. From top to bottom: 1) R$_{i,1}=$ R$_{i,2}=4.5$R$_*$ and $\Delta $R$_1=\Delta$R$_2 = 1$R$_*$; 2) R$_{i,1}=2.5$R$_*$, R$_{i,2}=5$R$_*$ and $\Delta$ R$_1=\Delta $R$_2 = 1$R$_*$; 3) R$_{i,1}=$ R$_{i,2}=4.5$R$_*$ and $\Delta $R$_1=0.5$R$_*$, $\Delta$R$_2 = 1$R$_*$; and 4) R$_{i,1}=2.5$R$_*$, R$_{i,2}=5$R$_*$ and $\Delta $R$_1=0.5$R$_*$, $\Delta$R$_2 = 1$R$_*$. Grey filled contours show the density of model predictions in 100 logarithmically spaced bins, with darker shading indicating a higher concentration of one-column models. Orange contours show the corresponding distribution for our CTTS sample, using linearly spaced bins. Different binning schemes were adopted to enhance readability, as the modeled dataset is much larger than the observed sample. Lines represent specific values of constant $\dot{M}_1$, while the overlaid scatter points show the resulting Balmer decrements for two-flow models, color-coded by $\dot{M}_2$. While not exhaustive, these examples highlight some of the effects that composite magnetospheres can have on Balmer decrements. Observed Balmer decrements can be explained with certain combinations of parameters in our two-column accretion model (see text.)}

    \label{fig:effect_2col}
\end{figure*}

\section{Balmer Decrements in multi-flow accretion}\label{decrements}

We calculated the Balmer decrements predicted by both the standard magnetospheric accretion model and our two-column model. 
The right panel of \autoref{fig:balmer_decrement} shows the ratios of H$\gamma$/H$\beta$ versus H$\alpha$/H$\beta$ for the two-column model. The filled contours represent the density of model predictions, computed in 100 logarithmically spaced bins, with darker shading indicating a higher concentration of models per bin.

In the left panel of \autoref{fig:balmer_decrement} we see that 
most one-column models cluster around H$\alpha$/H$\beta$ $\sim 4$ and H$\gamma$/H$\beta$ $\sim 0.3$. When compared to observations, it becomes clear that the underestimation of H$\gamma$ is not limited to a particular star or region. Instead, most of the one-column models predict H$\gamma$/H$\beta$ ratios that are too low to match the observations.

In contrast, the observed Balmer decrements 
can be reproduced with a subset of 
the proposed two-column model,
as shown in the right panel of \autoref{fig:balmer_decrement}. 

\medskip

\subsection{Exploring the parameter space in multi-flow accretion models}

To study the effects of multiple accretion columns on the Balmer 
decrements, we selected a representative model star and systematically varied the magnetospheric parameters. \autoref{fig:effect_2col} shows an example of this exercise using an M5 star, with parameters listed in Table \ref{table:parametros_modelos}.
Given the large number of possible parameter combinations, we constrain our analysis to a subset that allows us to explore how Balmer decrements respond to variations in characteristics of the two accretion columns.

For this test, we fix the coverage factors at $f_1 = 10^{-3}$ and $f_2 
= 3 \times 10^{-2}$, representing a low- and a high-coverage column, respectively. 

Each column corresponds to a different $T_{max,1}$, and each row corresponds to a different geometry.
We vary the accretion rates of both accretion flows and fix the maximum temperature, $T_{\mathrm{max}}$, to values representative of the range of accretion rates, following the relationship proposed by \citet{muzerolle2001}.

In each panel, circular markers represent values of $\dot{M}_2$, with values in the color bar on the right,
while each curve connecting the markers corresponds to a fixed $\dot{M}_1$. The number of curves shown varies between panels due to the constraint between T${_\mathrm{max}}$ and $\dot{M}$ (see Section \ref{individual_stars}). Orange contours indicate the region in the decrement space occupied by the majority of observed CTTS, while grey contours show the density of one-column model predictions in logarithmically spaced bins.

Overall, we find that \hgamma/\hbeta\ increases with the accretion rate of the flow with smaller coverage for most fixed values of the other column's accretion rate. In addition, two-column models that fall within the region that coincides with the observations tend to feature a large-coverage column with a relatively low accretion rate ($\dot{M}_2 < 10^{-9}$), besides a high $\dot{M}$ for the low coverage column (i.e., middle panel of second row).

Each row in \autoref{fig:effect_2col} illustrates the impact of changing the geometry of the accretion flows. These geometric changes produce varied effects, but we find that a broader range of accretion rates can reproduce the observed decrements when the two columns have significantly different geometries (second and bottom row). Although this analysis does not exhaust the full parameter space, it demonstrates that our two-column approach can reproduce observed Balmer decrements for specific parameter combinations.

\begin{figure}
    \centering
    \includegraphics[width=\linewidth]{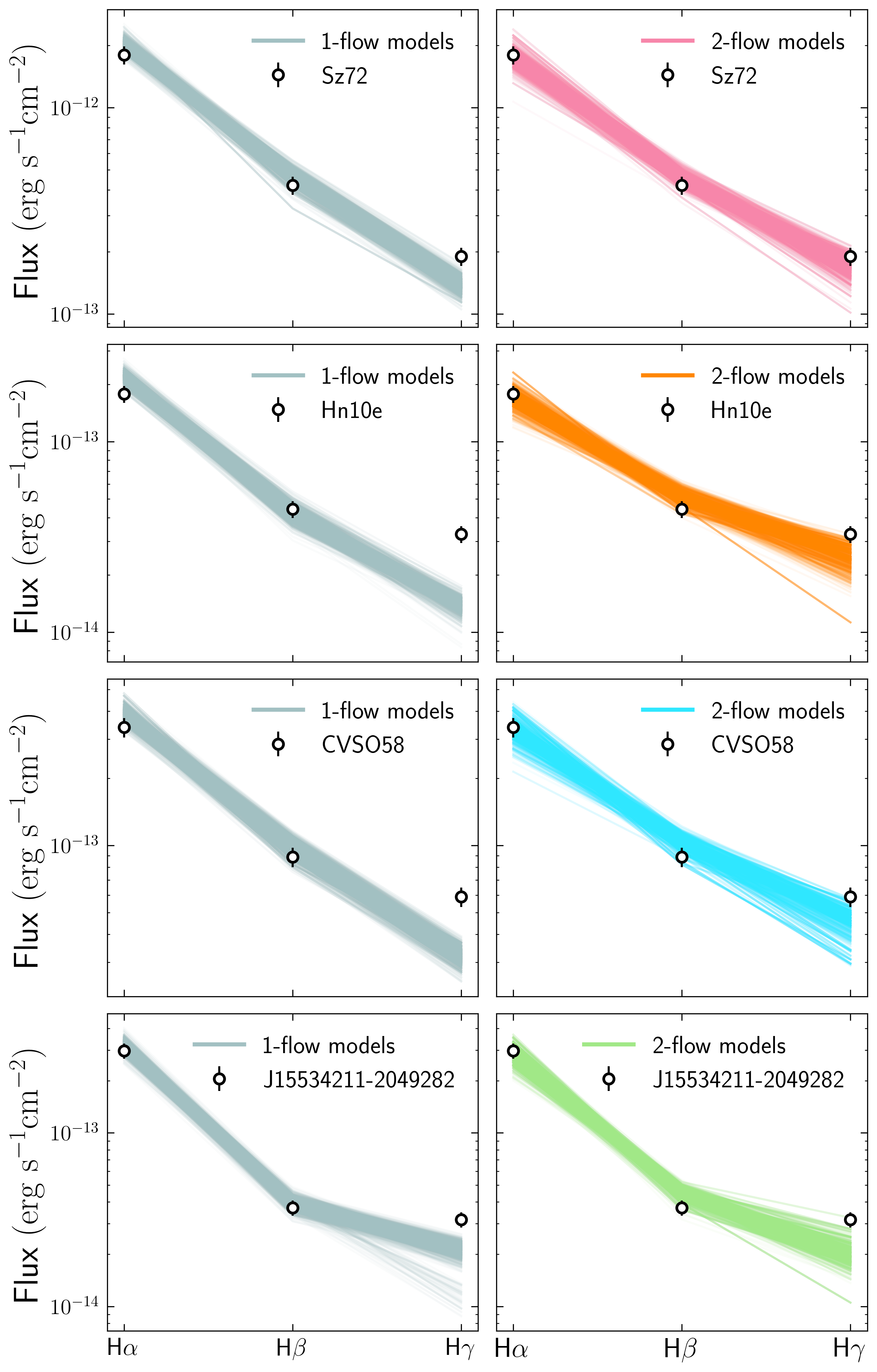}
    \caption{Example of fit to observed Balmer fluxes of two stars for each studied region using the one-column (left) and two-column (right) accretion model. Continuous lines connect the values of measured fluxes for \halpha, \hbeta, and \hgamma\ for 1000 models randomly extracted from the final, post-burn-in MCMC sample after 10,000 steps. Scatter points correspond to the dereddened measured fluxes from CTTS of each region.}  
    \label{fig:good_bad_fit}
\end{figure}

\section{Application to individual stars}\label{individual_stars}

In this section, we further explore the multi-flow models, by attempting to fit not only the line ratios but also the flux of the individual lines. To achieve this, we adopt a star-by-star approach and apply the two-column magnetospheric accretion model individually to each CTTS in our sample.

To estimate the composite model fluxes for each star, we compute magnetospheric accretion models using the model star from Table \ref{table:parametros_modelos} with the closest effective temperature to the observed CTTS. For that
model star, we run a grid of models that spans the parameter space defined in \autoref{table:parameter_space}, and  for every combination of model parameters, we calculate the flux of line $l$ using \autoref{eq:fmodcombined}. Before comparing the modeled fluxes to the observed ones, we scaled each model flux to the stellar distance $d_*$ and luminosity $L_*$ of the CTTS using

\begin{equation} \label{eq:scale_factor_mods}
    F_{l,\rm{comb}} = \frac{L_*}{L_{\rm{mod}}} \left(\frac{R_{\rm{mod}}}{d_*}\right)^2  F_{l,\rm{comb,unscaled}.}
\end{equation}

The chromospheric contribution has to be properly included in the flux of each line.
To do this, we used WTTS as templates for the chromospheric emission, as described in section \ref{chromospheric_emission}. We scaled the observed WTTS spectra to the CTTS using 

\begin{equation} \label{eq:scale_factor_WTTS}
    F_{\rm{WTTS}} =  F_{\rm{WTTS,obs}} \left(\frac{R_{\rm{CTTS}}}{d_{\rm{CTTS}}}\right)^2 \left(\frac{d_{\rm{WTTS}}}{R_{\rm{WTTS}}}\right)^2
\end{equation}

where $F_{\rm{WTTS}}$ is the flux of the WTTS scaled to the CTTS, $F_{\rm{WTTS,obs}}$ is the observed WTTS spectra, $R_{\rm{CTTS}}$ and $R_{\rm{WTTS}}$ are the radii of the CTTS and WTTS respectively, and $d_{\rm{CTTS}}$ and $d_{\rm{WTTS}}$ are their distances. If multiple WTTS of the same spectral type were available, we selected the one with the smallest H$\alpha$ FWHM. In cases where no exact spectral type match existed in our WTTS sample, we adopted the object with the closest spectral type. WTTS with missing line flux measurements were excluded from the sample. 

Finally, the model flux for each line $l$ is

\begin{equation} 
\label{eq:fluxmod}
    F_{l,\rm{mod}}  =  F_{l,\rm{comb}}  + F_{l,\rm{WTTS}} 
\end{equation}     

\subsection{MCMC}

We use the composite fluxes of \hgamma, \hbeta, and \halpha\ predicted by the magnetospheric accretion model to fit the line fluxes  in our sample of CTTS. 

We apply Bayesian statistics to infer the magnetospheric parameters of the model combinations that best reproduce the observed line fluxes. Our likelihood function is defined as:

\begin{equation}
\log (L) \propto -\frac{1}{2} \sum_l \frac{(F_{l,\text{obs}} - F_{l,\text{mod}})^2}{\sigma_l^2},
\end{equation}
where $F_{l,\text{obs}}$ is the observed flux of line $l$, $\sigma_l$ is its corresponding uncertainty, and $F_{l,\text{mod}}$ is the flux for that line predicted by our multi-flow model (eq. \ref{eq:fluxmod}). The index $l$ runs over the three Balmer lines under study.

We adopt the relationship between the maximum temperature and the mass accretion rate 
described by \citet{muzerolle2001} as a prior, excluding all models outside of this parameter space. The prior on \mdot\ is taken to be a Gaussian centered on the reported value (see \autoref{table:ctts_table}), with a width of $1\sigma$.

To sample the posterior probability distribution function (PPDF), we use the ensemble sampler implemented in the \texttt{emcee} package \citep{Foreman_Mackey_2013}. The Markov Chain Monte Carlo (MCMC) is run with 50 walkers for 10,000 steps, discarding the first 60\% of samples as burn-in. Given the complexity of the parameter distributions, we replace the default “stretch move” algorithm \citep{goodman-weare2010} with a combination of the more efficient “DEMove” \citep[][Differential Evolution]{Nelson_2014} 80\% of the steps, and "DESnookerMove" \citep[][Snooker algorithm with Differential Evolution]{DESnooker2008} the remaining 20\%. The final PPDFs for each star—referred to as the "best-fit models"—are obtained by sampling every 50 steps from the post-burn-in chain.

\begin{figure}
    \centering
    \includegraphics[width=\linewidth]{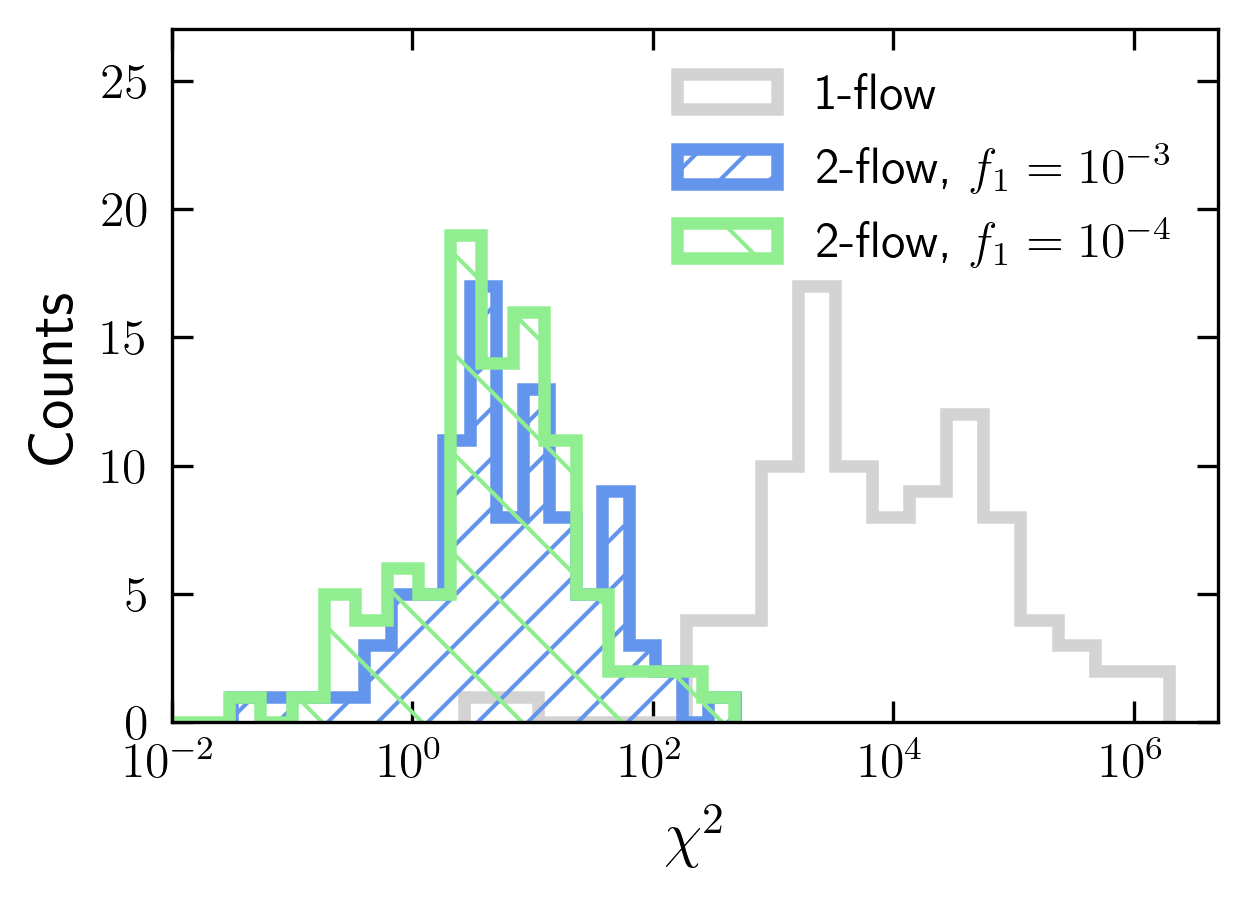}
    \caption{Distribution of minimum $\chi^2$ values for the best-fit models using one-flow (grey), two-flows with $f_1 = 10^{-3}$ (blue), and two-flows with $f_1 = 10^{-4}$ (green) for our sample. We observe no significant difference between the two choices of $f_1$ for the two-flow model, but there is a significant improvement of the fit compared to one-flow models.}
    \label{fig:chi2_comparison}
\end{figure}

\begin{figure}
    \centering
    \includegraphics[width=0.9\linewidth]{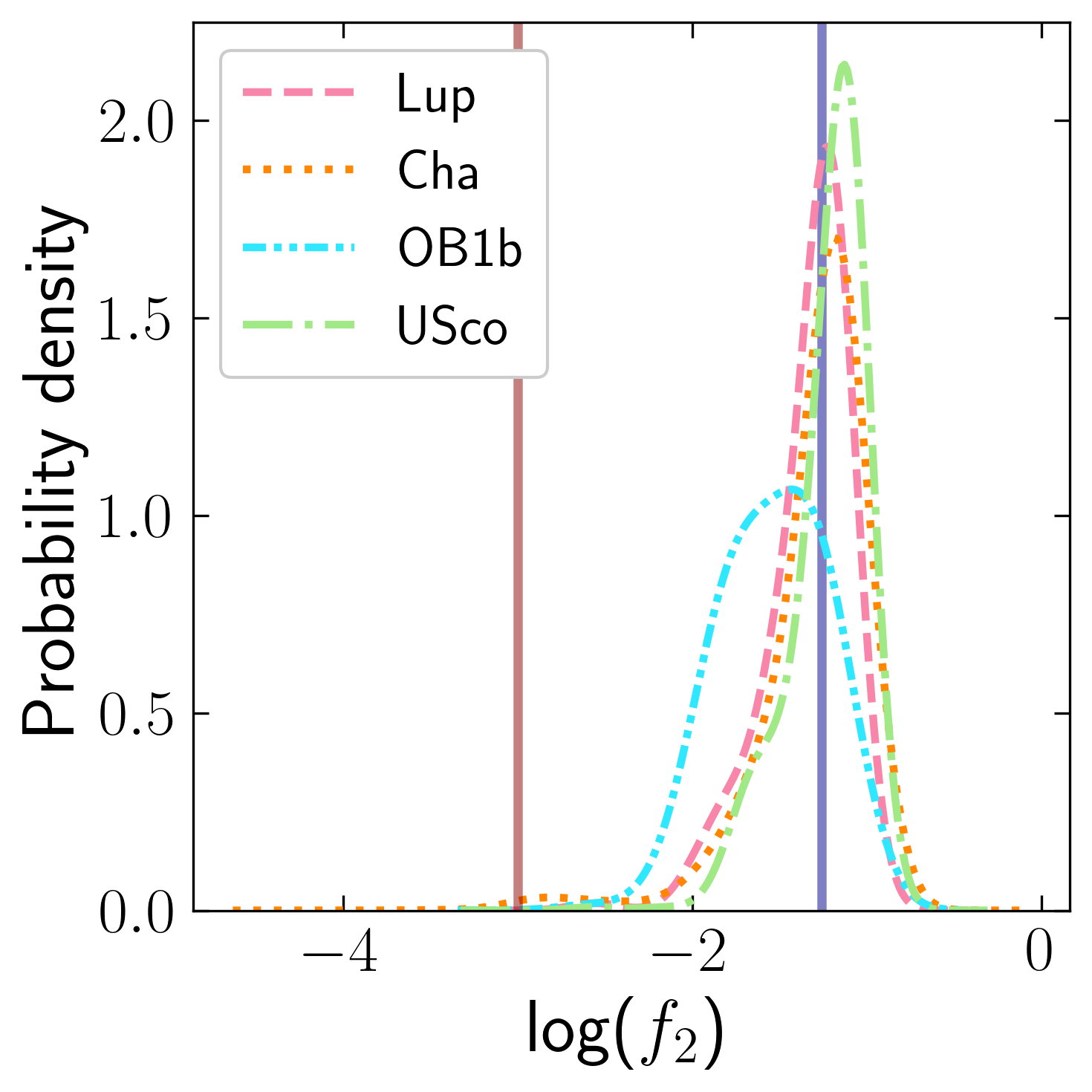}
    \caption{Distribution of $f_2$ for the best-fit models. Each curve corresponds to a different star-forming region. The red line indicates the fixed value adopted for one 
    flow, $f_1 = 10^{-3}$, while the blue line marks the weighted mean of the $f_2$ distribution across all stars in the sample. To emphasize models with well-constrained values, each $f_2$ value is weighted by the squared inverse of its associated uncertainty.}
    \label{fig:result_logf2}
\end{figure}

\begin{figure*}
    \centering
    \includegraphics[width=1\linewidth]{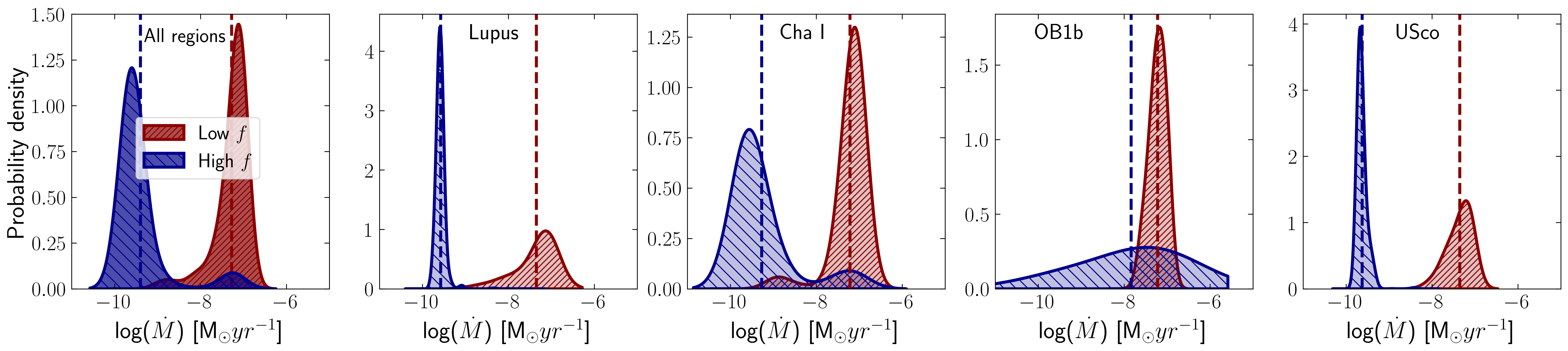}
    \caption{Distribution of accretion rates for each magnetospheric 
    flow. The filled curves show the Kernel Density Estimation (KDE) of the distributions, with blue representing the high-coverage flow and red the low-coverage flow. Each point corresponds to the median accretion rate from the posterior probability distribution obtained via the MCMC analysis for a given star. Vertical dashed lines indicate the weighted median of each distribution. The left panel shows results for the full sample across all regions, while the remaining panels present the distributions separated by star-forming region (SFR), increasing in age towards the right.}
    \label{fig:mdot_result}
\end{figure*}

\subsection{Line fluxes}

As shown in \autoref{decrements}, the proposed two-flow model successfully reproduces the observed Balmer decrements (see right panel of \autoref{fig:balmer_decrement}). For most of our sample, this composite model also provides a better match to the total emission fluxes of individual stars than the standard one-flow model, as illustrated in \autoref{fig:good_bad_fit}, where we show the  \halpha, \hbeta, and \hgamma\ fluxes  predicted by our model for a star of each region. More broadly, \autoref{fig:chi2_comparison} compares the $\chi^2$ values obtained for the two-flow models with different choices of $f_1$ to those of the one-flow model, where $\chi^2$ is defined as
\begin{equation}
\chi^2 = \sum_l \frac{(F_{l,\text{obs}} - F_{l,\text{mod}})^2}{\sigma_{l,\text{obs}}^2},
\end{equation}
where $l$ identifies each of the studied lines. With this definition, lower $\chi^2$ values correspond to models that better reproduce the three line fluxes simultaneously. We find no substantial differences among the fits for different values of $f_1$, but a significant improvement relative to the $\chi^2$ values of the one-flow models. The remaining fits to the total emission fluxes for the full sample are shown in the online repository \citep{zenodo_repo}.

\subsection{Objects with high chromospheric contribution}

We approximate the chromospheric contribution in our sample of CTTS using WTTS templates of the same spectral type. However, chromospheric emission is highly uncertain — it can vary over time and between individual stars. To avoid introducing this uncertainty into the statistical analysis of the best-fit models (see \autoref{accretion_res}, \autoref{geo_res}), we exclude from the analysis all objects for which the chromospheric contribution exceeds 30\% in at least one of the emission lines of interest. This includes objects where the WTTS template is not representative of the star's actual chromospheric activity —leading to poor fits — as well as objects for which, despite successful fits with our multiflow magnetospheric model, the emission is dominated by chromospheric processes and could bias the inferred parameter distributions.

\subsection{Coverage factor and accretion rate} \label{accretion_res}

\begin{figure}
    \centering
    \includegraphics[width=\linewidth]{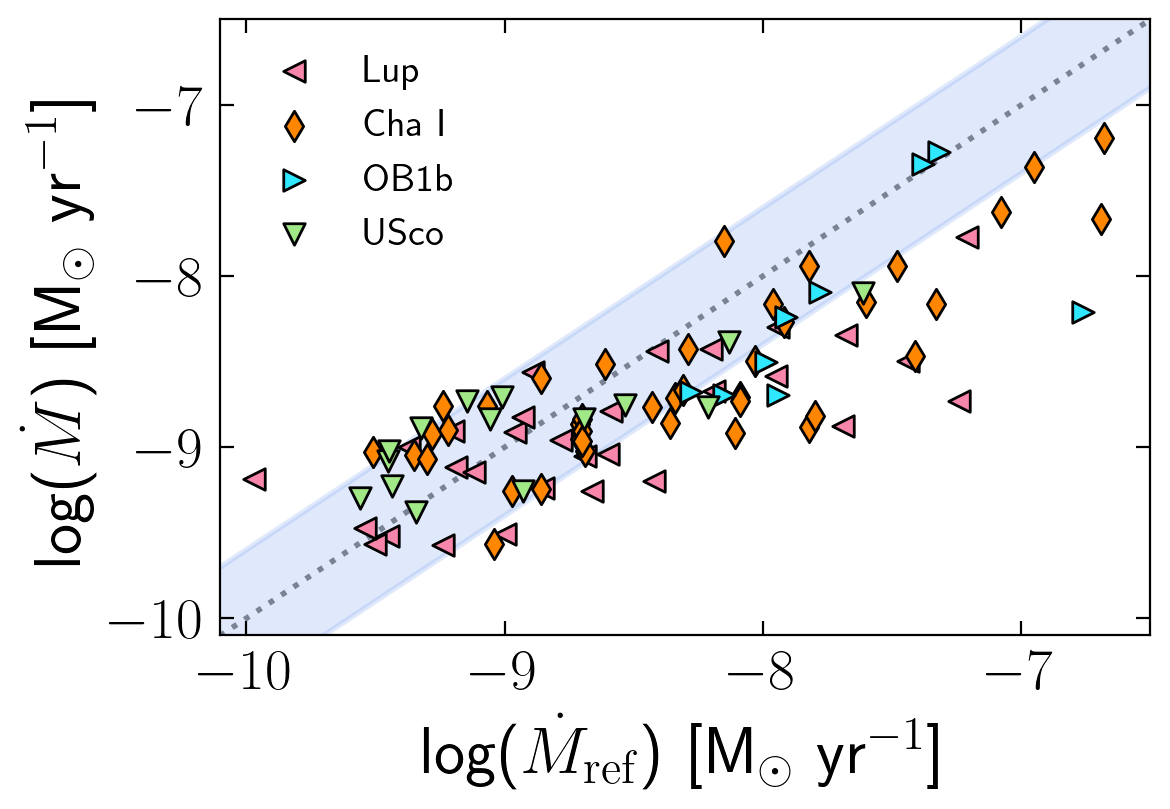}
    \caption{Comparison of the mass accretion rates derived in this work using the two-flow magnetospheric accretion model with values reported in the literature (see \autoref{table:ctts_table}). The dotted gray line denotes the identity relation, indicating a one-to-one correspondence between our results and the literature values. The blue shaded region represents a 0.4 dex deviation from this relation, corresponding to the expected dispersion of accretion rates due to variability \citep{manara2023}.}
    \label{fig:my_mdot_vs_ref}
\end{figure}

In this section, we discuss the distribution of accretion rates and coverage factors ($f_i$) of the best-fit models to the objects in our sample. For each star and flow, we extracted the median value of the parameter of interest and its uncertainty from the posterior distribution sampled by the MCMC, where the uncertainty is defined as the range between the 16th and 84th percentiles. To reduce the effect of poorly constrained parameters, the median values extracted for individual stars were weighted by their inverse uncertainties when constructing the final distributions. \autoref{fig:result_logf2} shows the distribution of coverage factors found in our sample. Note that this factor is fixed for one column ($f_1=10^{-3}$, red), while the other is a free parameter. We obtain similar distributions across star-forming regions, with a weighted mean of $\overline{f_2}=10^{-1.2}$ (blue).

The distributions in \autoref{fig:result_logf2} indicate that the best-fit models tend to have one flow that occupies $\frac{\overline{f_2}}{\overline{f_2}+f_1}\sim 98\%$ of the emitting area, while the other covers the remaining $\frac{f_1}{\overline{f_2}+f_1}\sim 2\%$. Hereafter, we will refer to them as high- and low-coverage flows, respectively. 

\autoref{fig:mdot_result} presents the distribution of accretion rates corresponding to the best-fit models for the flows with low-coverage (red) and high-coverage (blue). The left panel shows the full sample, while the remaining panels display results
for each star-forming region separately, which can be used as indicative of age. We find no indication of a trend with age, although a larger sample of star-forming regions would be needed to confirm this result. From this, we find that the low-coverage flows of the best-fit models tend to have higher accretion rates than the flows with high-coverage. Ori Ob1b is the only region where this separation does not hold and both flows tend to have high mass accretion rates. However, 
inspection of Figure \ref{fig:params_CTTS} shows that the OB1b sample lacks stars with low accretion rates and is the smallest of the samples, making direct comparisons with other regions less robust.

We calculate a composite accretion rate from our two-flow model, and then compare it with the reference values from the literature. In \autoref{fig:my_mdot_vs_ref}, we show the relationship between the mass accretion rates calculated in this work and those extracted from the literature. The line of slope 1 is shown as a dotted line while the black dashed line is the linear fit to our results, with a grey shaded region representing the 95\% confidence interval. We find that the accretion rates of the two-flow models are comparable to those derived in the literature from measurements of the excess emission over the photosphere except for high mass accretion rates. This may be due to inherent limitations of the two-flow models, in that more complex geometries are required to describe the accretion regions of the highest accretors.
\medskip

\subsection{Accretion geometry}\label{geo_res}

\begin{figure}
    \centering
    \includegraphics[width=1\linewidth]{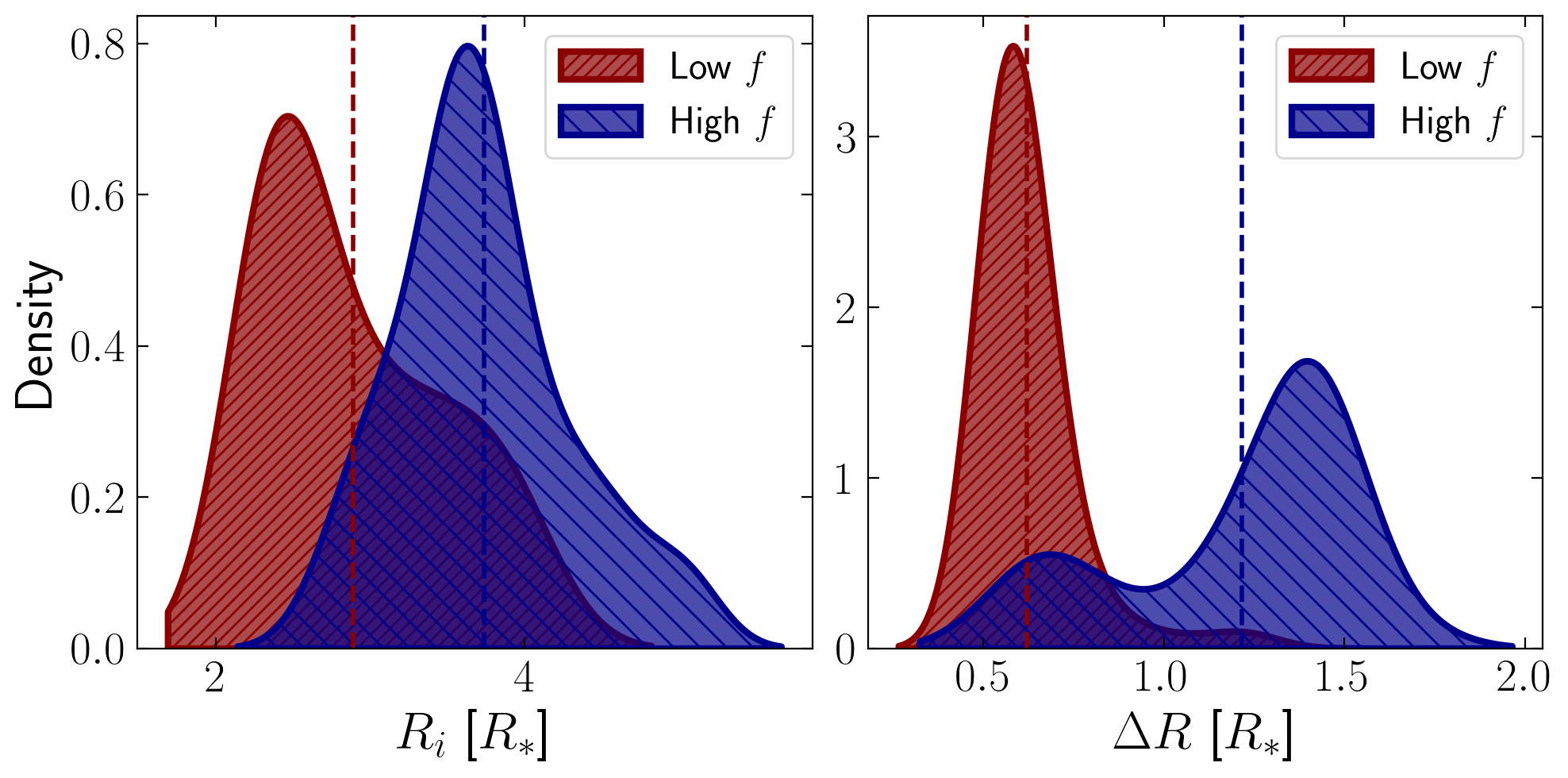}
    \caption{Trends in magnetospheric geometry. The filled curves show the KDE of the parameter distributions. Each point represents the median of the posterior probability distribution obtained from the MCMC analysis for a given star and parameter. Vertical dashed lines indicate the weighted median of each final distribution across the full sample.\textit{ Left:} Distribution of inner magnetospheric radius, $R_i$. Low-density flows (blue) show a weighted mean radius of $\overline{R_{i,2}} = 3.7 R_*$, while high-density flows (red) have $\overline{R_{i,1}} = 2.9 R_*$. \textit{Right:} Distribution of flow thickness, $\Delta R$. The weighted mean is $\overline{\Delta R_2} = 1.2 R_*$ for the high-coverage column (blue) and $\overline{\Delta R_{1}} = 0.6 R_*$ for the low-coverage column (red).}
    \label{fig:geometry_results}
\end{figure}

\begin{figure}
    \centering
    \includegraphics[width=1\linewidth]{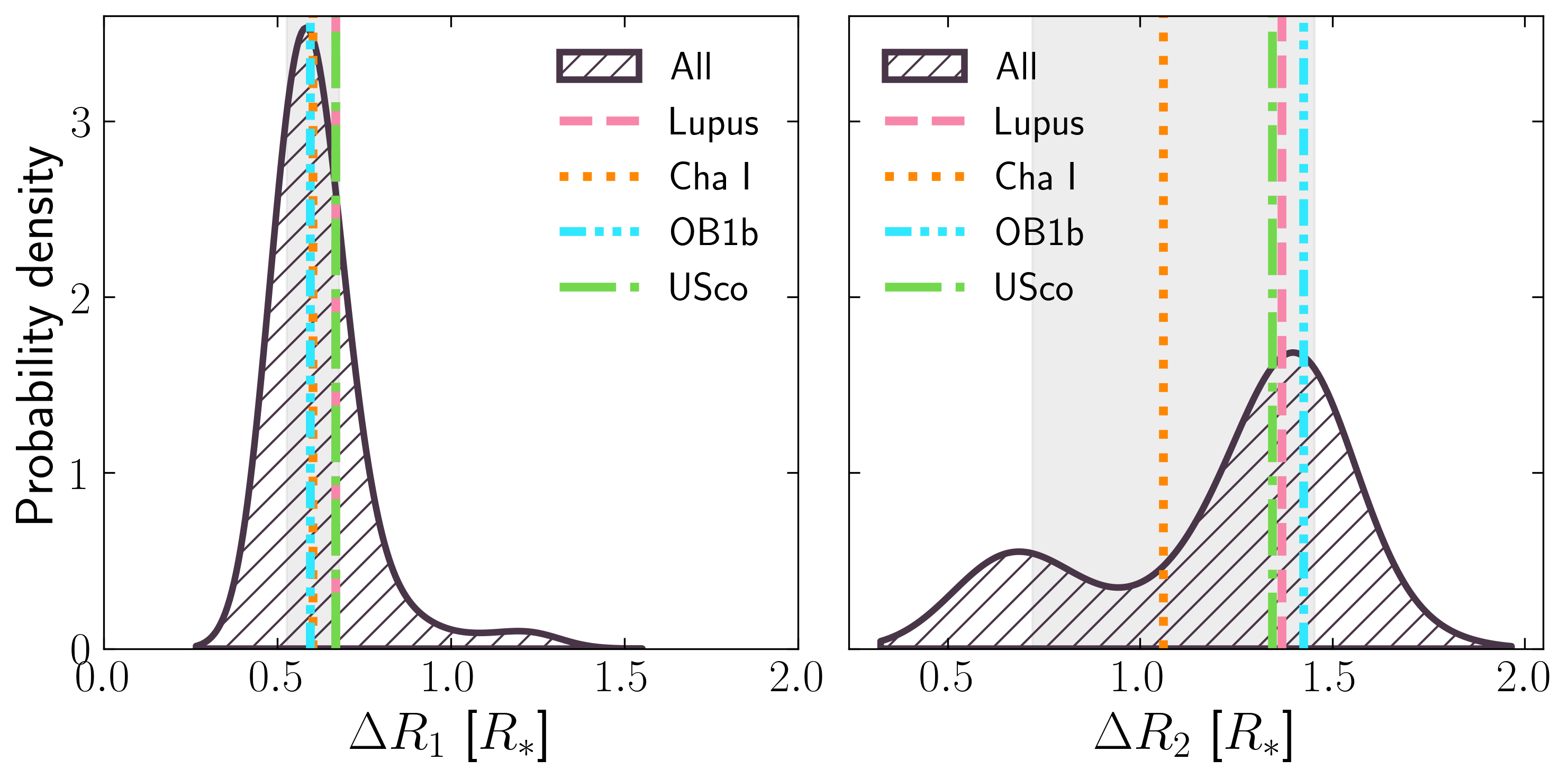}
    \caption{Trends in magnetospheric geometry by star-forming region. The filled curves show the Kernel Density Estimation (KDE) of the parameter distributions. Each point represents the median of the posterior probability distribution obtained from the MCMC analysis for an individual star and parameter. Vertical dashed lines indicate the weighted median of each distribution for each region. The shaded area shows the range between the 16th and 84th percentiles of the combined distribution across all regions.}
    \label{fig:geo_by_region}
\end{figure}

The distribution of the best-fit models in the parameter space shows certain trends for the geometry of the two accretion flows. \autoref{fig:geometry_results} shows the distribution of truncation radii ($R_i$) and column thickness ($\Delta R$). Similarly to the results in \autoref{accretion_res}, we extracted the median value for these parameters and their uncertainty from the posterior distribution sampled by the MCMC, where the uncertainty is defined as the range between the 16th and 84th percentiles. The median values for individual stars were weighted by their inverse squared uncertainties when constructing the final distributions. Using this approach, we find that the high-coverage column (blue) has a typical truncation radius of $\overline{R_{i,2}} = 3.7 R_*$ across all regions. However, for the low-coverage column (red), we find that, in general, smaller truncation radii are preferred, resulting in a weighted mean of $\overline{R_{i,1}} = 2.9 R_*$. A similar trend can be observed in the column thickness, where high-coverage columns tend to have thicker columns, $\overline{\Delta R} = 1.2 R_*$, while low-coverage columns tend to be thinner, $\overline{\Delta R} = 0.6 R_*$. We do not find a significant difference in these values when we group stars by star-forming region, which indicates no trend with age as shown in \autoref{fig:geo_by_region}. We obtain that the weighted mean value for $\Delta R$ and $R_i$ for each region, falls within the 16th and 84th percentiles of the distribution including all objects.

Combining the results of sections \ref{accretion_res} and \ref{geo_res}, we find that the models that best fit the Balmer line fluxes can be described by two distinct but coexisting types of accretion flows: a  geometrically small, compact, flow with small inner radius and width and a high mass accretion rate, covering a low fraction of the emitting area, and a larger, more extended flow, with lower mass accretion rate and a higher surface coverage. 
\medskip

\section{Discussion} \label{Discussion}

\subsection{Approximating a inhomogeneous magnetosphere}

Magnetic fields in classical T Tauri stars are inherently complex, and non-axisymmetric accretion is expected to be common \citep{romanova2003}. This complexity implies that the traditional model of uniform, homogeneous accretion flows may be insufficient to extract some valuable information about the characteristics of these accretion columns. In this study, we approximate this complex behavior by adopting a two-flow magnetosphere model. Despite its simplicity, this approach reproduces the observed Balmer line emission fluxes more accurately than a 
single-flow model. Although it remains a coarse representation of the true geometry of the stellar magnetic fields and the infalling matter, it offers valuable insights into the accretion processes in individual stars and across different star-forming regions. 

With our simple approach, we obtain results consistent with complex MHD simulations. \citet{romanova2003} suggest a connection between the complicated spectral variability patterns in CTTS and the structure of the magnetospheric flows. They find that the geometry of the magnetospheric flows depends on the density of the flow, with fewer and narrow high density columns.
This is consistent with our results, since the flows with higher mass accretion rate - and thus high density, since they tend to be compact - occupy a smaller fraction of the emitting area.

Similarly, \citet{zhaohuan2024} find complex density structures in the accretion flows, using 3-D MHD simulations. They 
find filaments with densities up to 3 orders of magnitude higher than the background, and accretion streams forming simultaneously and independently. We suggest that the high-density columns we obtain in our simplified model may be an approximation of these structures. 

In the long term, the goal is to find a way to relate our simple approach to the results of numerical simulations, to be able to better describe the complexity of the line emission from the magnetospheric infall regions, and address questions such as line variability. 

\subsection{\hgamma\ as a tracer of high-accretion columns}
\begin{figure}
    \centering
    \includegraphics[width=0.9\linewidth]{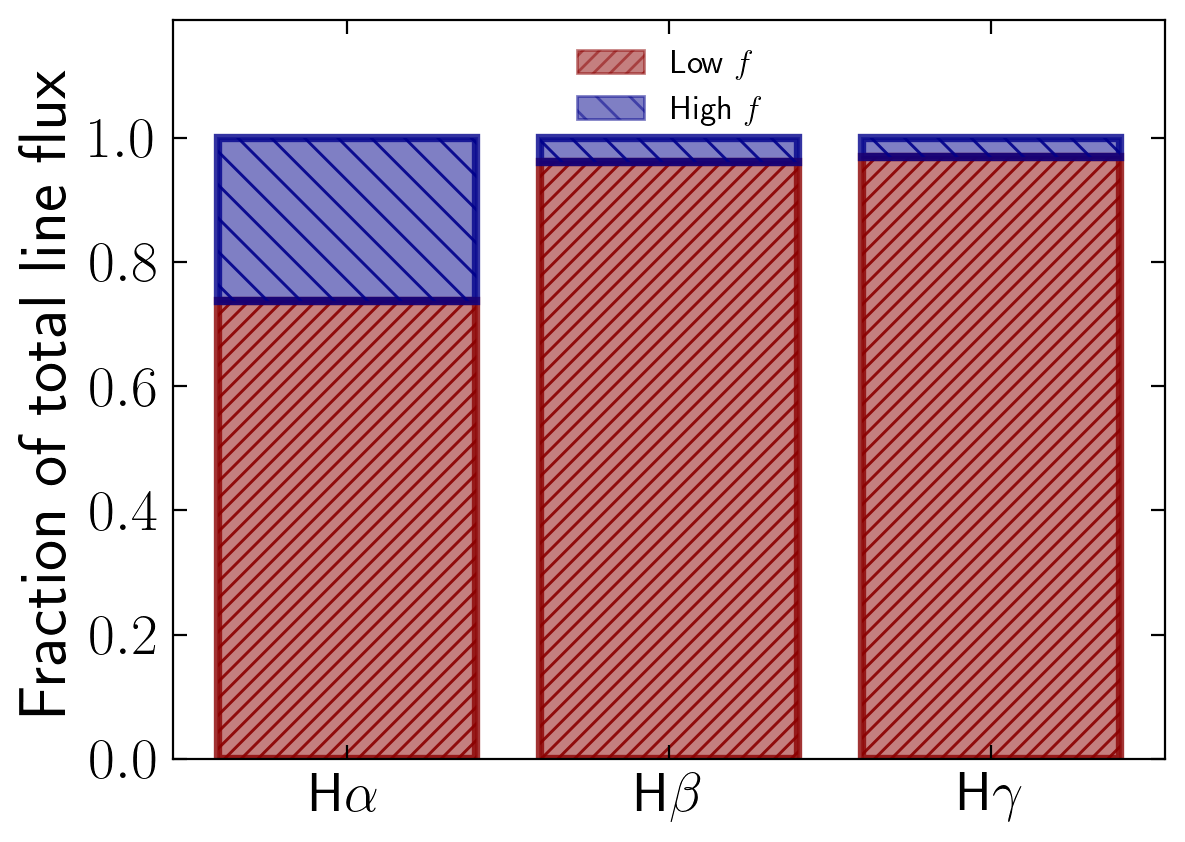}
    \caption{Fraction of flux emitted by each column. This figure shows the typical fraction of the total emission contributed by each magnetospheric column—low-coverage (red) and high-coverage (blue)—for each emission line. For each star, we calculate the mean fraction of the total emission that each column produced. The final values plotted represent the mean of these fractions across all stars, weighted by the inverse square of the standard deviation of the flux distribution from the best-fit models.}
    \label{fig:fraction_by_col}
\end{figure}

\begin{figure*}
    \centering
    \includegraphics[width=\linewidth]{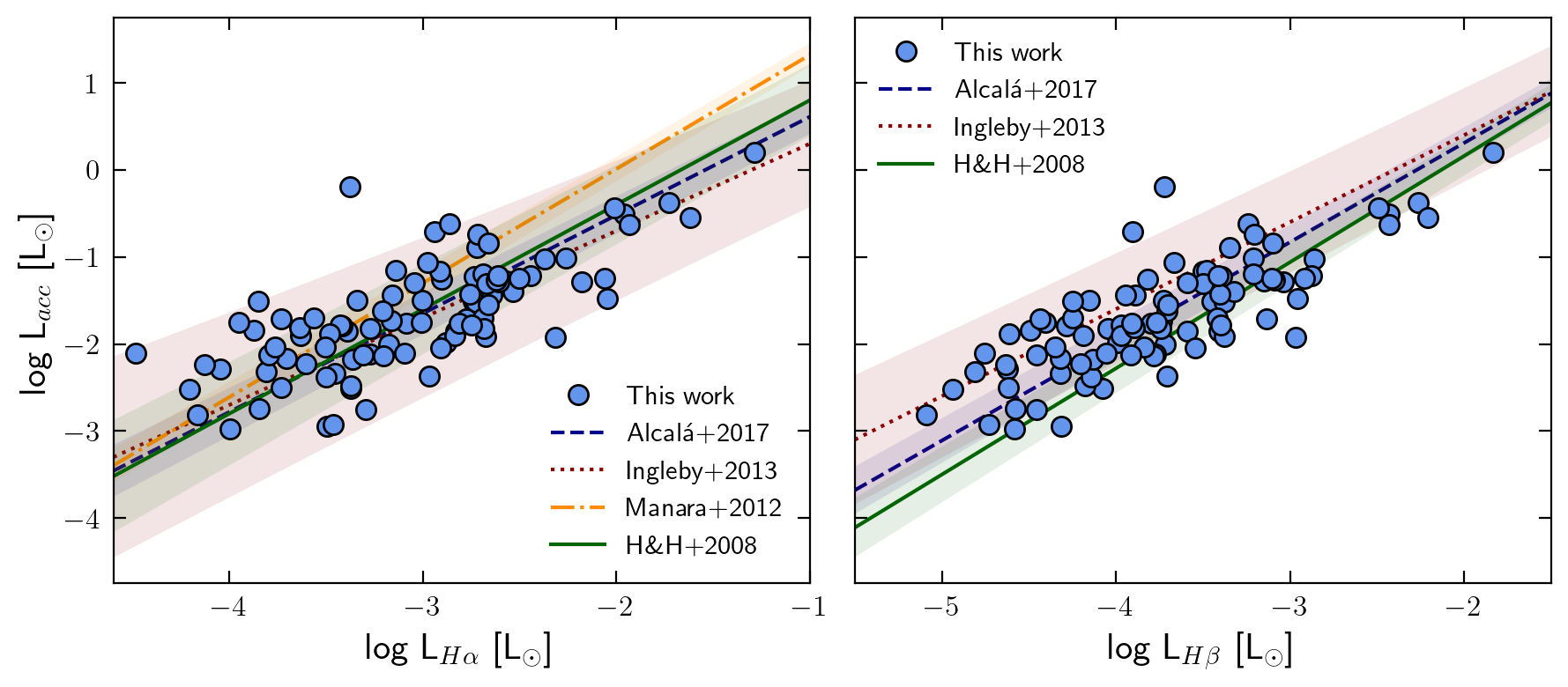}
    \caption{ Relationship between accretion and line luminosity predicted by the two-column accretion model. The accretion luminosity was calculated from the composite accretion rates of the best-fit models for each object (scattered points) using \autoref{eq:lacc_v_lhalpha}. The lines in both panels represent relationships reported in the literature, while the shaded regions indicate their associated uncertainties. \textit{Left:} Accretion luminosity versus \halpha: \citet{ingleby2013} (red dotted line), \citet{alcala2017} (blue dashed line), \citet{manara2012} (orange dot-dashed line), and \citet{HerczegHillenbrand2008} (H\&H+2008) (solid green line). \textit{Right:} Accretion luminosity versus \hbeta: \citet{ingleby2013} (red dotted line), \citet{alcala2017} (blue dashed line), and \citet{HerczegHillenbrand2008} (H\&H+2008) (solid green line).
    The composite accretion rates derived from the two-column model for both lines are consistent with these empirical relationships within the estimated uncertainties.}
    \label{fig:lacc_v_lalpha}
\end{figure*}

Our analysis allows us to consider the contribution of each flow to the total flux of a given line, using our extended stellar sample. To do so, for each star in the sample, we compute the median fraction of the total emission that is emitted by each flow. We then average these fractions across all stars, weighting by the standard deviation of the inverse square of the flux distribution of the best-fit models. 
\autoref{fig:fraction_by_col} shows the fractions of the flux emitted by the low and high surface coverage flows resulting from this global analysis. 
It can be seen that most of the \hgamma\  emission originates in the low-coverage, high-mass accretion rate flow. This suggests that \hgamma\ could serve as a tracer of these denser flows; additionally, it indicates that  previous attempts to fit fluxes of Balmer lines were underestimating \hgamma\ emission due to a lack of a higher accretion rate component. This is consistent with the results of \citet{wendeborn2024}, who use the magnetospheric accretion model to study the variability of \halpha, \hbeta, and \hgamma\ individually in four CTTS and consistently find that best-fit models to \hgamma\ have a higher accretion rate.

As the variability of spectral lines depends on density, temperature, and velocity distributions along the line of sight, multiple accretion columns rotating with the star may produce observable variability in Balmer line profiles—even in the absence of changes in the total accretion rate \citep{wendeborn2024}. Our results support this interpretation by showing that distinct accretion streams can contribute very differently to the line emission.

\subsection{Relation between line luminosity and accretion luminosity}

Correlations between the luminosity of emission lines, such as H$\alpha$ or Br$\gamma$, 
and accretion luminosity have been extensively used to find the accretion luminosity and thus the mass accretion rate in stars for which 
direct measurements of the excess energy due to accretion is not available. This is mostly the case for young, embedded objects where the UV excess is not accesible \citep{muzerolle1998b}, or more generally when short wavelength observations are not available.

These correlations are usually calibrated with stars that have both lines luminosities and accretion luminosities determined from measurements of excess emission over the photosphere using slabs or accretion shock models \citep[cf.][]{calvet2004,ingleby2013}, ideally simultaneous. But so far, magnetospheric models have not been able to explain these empirical correlations \citep{muzerolle2001}. We can follow the same treatment as in section \ref{accretion_res} to test the two-flow model. Specifically, we use the calculated composite accretion rates from the best-fit models of the sample to obtain the accretion luminosity $L_{\text{acc}}$ of each star

\begin{equation} \label{eq:lacc_v_lhalpha}
    L_{\text{acc}} = \frac{G M_{*} \dot{M}}{R_*}
\end{equation}


\autoref{fig:lacc_v_lalpha} shows the accretion luminosity vs. the \halpha\ luminosity from the two-flow model for all stars in the sample. The lines in \autoref{fig:lacc_v_lalpha} are empirical relations found in \citet{HerczegHillenbrand2008}, \citet{manara2012}, \citet{ingleby2013}, and \citet{alcala2017}, in which $L_{\text{acc}}$ was determined by fitting accretion shock models to HST-optical data.

\autoref{fig:lacc_v_lalpha} shows that the two-flow model can explain the empirical relations between line and accretion luminosities. \citet{muzerolle2001} had already suggested that departures from an axi-symmetric magnetosphetic model were needed to explain the empirical correlations. Here, we confirm this interpretation, by using the two-flow model, a proxy for a complex magnetospheric flow, to successfully model the empirical correlation between  $L_{\text{acc}}$ and $L_{H\alpha}$. 


\section{Summary and Conclusions}

We have applied a two-flow magnetospheric model to calculate the fluxes of {\halpha}, {\hbeta}, and \hgamma\ in a sample of 139 stars 
taken from the Ori OB1b subassociation and from the Lupus, Chamaeleon I, and Upper Sco star-forming regions. We find:

\begin{enumerate}

\item Even a simplified two-column model can capture key features of the complex accretion geometry in classical T Tauri stars, specifically the Balmer decrement and individual line fluxes.

\item The models that best fit the observed Balmer line fluxes 
consist
of two distinct but coexisting flows: a  compact flow with a small inner magnetospheric radius and width and a high mass accretion rate, covering a few percent of the emitting area, and a  more extended flow, with lower mass accretion rate
encompassing the rest of the emitting area

\item The anticorrelation between surface coverage and mass accretion rate - or density - of the flows is consistent in general terms with predictions of numerical simulations \citep{romanova2003,zhaohuan2024}.

\item While the compact, high accretion rate flow contributes the most to all lines, it dominates in  lines that require denser and hotter conditions to form, such as {\hgamma}

\item The two-flow model can reproduce the empirical correlation between the luminosity in \halpha\ and the accretion luminosity

\end{enumerate}

\bigskip
\bigskip

\section*{Acknowledgements}

We thank the reviewer for their insightful feedback, which has helped improve this manuscript. N.P. thanks all members of the research group "Models and Observations of Disk Evolution in Latin America" (MODELA), Prof. Catherine Espaillat and members of her research group: Caeley Pittman, Máire Volz, Luisa Zamudio-Ruvalcaba and Chunhua "Charlie" Qi, Prof. Alberto Patiño, Katya Gozman, Sujay Shankar and Stefan Arseneau, for valuable discussions and guidance throughout this project. N.C. would like to acknowledge  the ESO Scientific Visitor Programme for providing support for a visit in which many ideas for this paper were developed.  

We are grateful to the developers of the open-source software packages that enabled our analysis.

This work was partially supported by the NASA XRP 80NSSC2K0151 grant.

\software{\texttt{Astropy} \citep{astropy:2013, astropy:2018, astropy:2022}, \texttt{emcee} \citep{Foreman_Mackey_2013}, \texttt{SciPy} \citep{scipy2020}, \texttt{NumPy} \citep{numpy2020}}

\pagebreak

\bibliography{references.bib}{}
\bibliographystyle{aasjournal}

\appendix
\restartappendixnumbering

\section{Tables}

\input{table_ctts_updated}
\input{table_wtts}

\end{document}

%% file: table_ctts_updated.tex
\startlongtable
\begin{deluxetable*}{ccccccccc}
\tablecaption{Stellar parameters of the CTTS sample}\label{table:ctts_table}
\tablewidth{0pt}
\tabletypesize{\footnotesize}
\tablehead{
\colhead{Object} & \colhead{$SpT$} & \colhead{\teff\ [K]} & 
\colhead{$A_V$\ [mag]} & \colhead{$d$\ [pc]} &  
\colhead{$L_\star$\ [$L_\odot$]} & 
\colhead{$R_\star$\ [$R_\odot$]} & 
\colhead{$M_\star$\ [$M_\odot$]} & 
\colhead{log($\dot{M}$)\ [$M_\odot$/yr]}}
\startdata
\cutinhead{\hspace{2cm}Chamaleon I}
CHX18N &	             K2.0 & 4900 &		     0.8 & 	160 &		    1.03 &		         1.41		     &  1.25 &		     -8.09	\\
CHX18N &	             K2.0 & 4900 &		     0.8 & 	160 &		    1.03 &		         1.41		     &  1.25 &		     -8.09	\\
CRCha &	                 K0.0 & 5110 &		     1.3 & 	160 &		    3.26 &		         2.3		     &  1.77 &		     -8.71	\\
CSCha &	                 K2.0 & 4900 &		     0.8 & 	160 &		    1.45 &		         1.67		     &  1.4	 &	         -8.29	\\
CTChaA &	             K5.0 & 4350 &		     2.4 & 	160 &		    1.5	 &	             2.16		     &  0.98 &		     -6.69	\\
CWCha &	                 M0.5 & 3780 &		     2.1 & 	160 &		    0.18 &		         0.989		     &  0.59 &		     -8.03	\\
ESO-Ha-562 &	         M1.0 & 3705 &		     3.4 & 	160 &		    0.12 &		         0.841		     &  0.56 &		     -9.24	\\
Hn10e &	                 M3.0 & 3415 &		     2.1 & 	160 &		    0.06 &		         0.7		     &  0.34 &		     -9.51	\\
Hn5 &	                 M5.0 & 3125 &		     0.0 & 	160 &		    0.05 &		         0.763		     &  0.16 &		     -9.28	\\
Hn13 &	                 M6.5 & 2935 &		     1.3 & 	160 &		    0.13 &		         1.39		     &  0.12 &		     -9.57	\\
Hn17 &	                 M4.5 & 3200 &		     0.4 & 	160 &		    0.11 &		         1.08		     &  0.2	 &	         -9.71	\\
Hn18 &	                 M4.0 & 3270 &		     0.8 & 	160 &		    0.11 &		         1.03		     &  0.24 &		     -9.81	\\
Hn21W &	                 M4.5 & 3200 &		     2.2 & 	160 &		    0.12 &		         1.13		     &  0.2	 &	         -9.04	\\
ISO-ChaI-282 &	         M5.5 & 3060 &		     2.8 & 	160 &		    0.07 &		         0.941		     &  0.14 &		     -9.89	\\
J11085367-7521359 &	     M1.0 & 3705 &		     1.5 & 	160 &		    0.19 &		         1.06		     &  0.51 &		     -8.15	\\
J11183572-7935548 &	     M5.0 & 3125 &		     0.0 & 	160 &		    0.26 &		         1.74		     &  0.19 &		     -8.95	\\
J11432669-7804454 &	     M5.5 & 3060 &		     0.4 & 	160 &		    0.09 &		         1.07		     &  0.14 &		     -8.71	\\
SzCha &	                 K2.0 & 4900 &		     1.3 & 	160 &		    1.17 &		         1.5		     &  1.31 &		     -7.82	\\
Sz18 &	                 M2.0 & 3560 &		     1.3 & 	160 &		    0.26 &		         1.34		     &  0.38 &		     -8.7	\\
Sz22 &	                 K5.0 & 4350 &		     3.2 & 	160 &		    0.51 &		         1.26		     &  1.01 &		     -8.34	\\
Sz27 &	                 K7.0 & 4060 &		     2.9 & 	160 &		    0.33 &		         1.16		     &  0.8	 &	         -8.86	\\
Sz32 &	                 K7.0 & 4060 &		     4.3 & 	160 &		    0.48 &		         1.4		     &  0.78 &		     -7.08	\\
Sz33 &	                 M1.0 & 3705 &		     1.8 & 	160 &		    0.11 &		         0.805		     &  0.56 &		     -9.35	\\
Sz37 &	                 M2.0 & 3560 &		     2.7 & 	160 &		    0.15 &		         1.02		     &  0.41 &		     -7.82	\\
Sz45 &	                 M0.5 & 3780 &		     0.7 & 	160 &		    0.42 &		         1.51		     &  0.51 &		     -8.09	\\
T10 &	                 M4.0 & 3270 &		     1.1 & 	160 &		    0.1	 &	             0.985		     &  0.23 &		     -9.22	\\
T12 &	                 M4.5 & 3200 &		     0.8 & 	160 &		    0.15 &		         1.26		     &  0.19 &		     -8.7	\\
T16 &	                 M3.0 & 3415 &		     4.9 & 	160 &		    0.28 &		         1.51		     &  0.29 &		     -7.8	\\
T23 &	                 M4.5 & 3200 &		     1.7 & 	160 &		    0.32 &		         1.84		     &  0.21 &		     -8.11	\\
T24 &	                 M0.0 & 3850 &		     1.5 & 	160 &		    0.4	 &	             1.42		     &  0.58 &		     -8.49	\\
T27 &	                 M3.0 & 3415 &		     1.2 & 	160 &		    0.34 &		         1.67		     &  0.29 &		     -8.36	\\
T28 &	                 M1.0 & 3705 &		     2.8 & 	160 &		    0.3	 &	             1.33		     &  0.48 &		     -7.92	\\
T3 &	                 K7.0 & 4060 &		     2.6 & 	160 &		    0.18 &		         0.858		     &  0.77 &		     -8.61	\\
T3B &	                 M3.0 & 3415 &		     1.3 & 	160 &		    0.19 &		         1.25		     &  0.29 &		     -8.43	\\
T30 &	                 M3.0 & 3415 &		     3.8 & 	160 &		    0.16 &		         1.14		     &  0.3	 &	         -8.31	\\
T33A &	                 K0.0 & 5110 &		     2.5 & 	160 &		    1.26 &		         1.43		     &  1.26 &		     -8.97	\\
T33B &	                 K0.0 & 5110 &		     2.7 & 	160 &		    0.69 &		         1.06		     &  1.0	 &	         -8.69	\\
T38 &	                 M0.5 & 3780 &		     1.9 & 	160 &		    0.13 &		         0.841		     &  0.63 &		     -9.3	\\
T4 &	                 K7.0 & 4060 &		     0.5 & 	160 &		    0.43 &		         1.33		     &  0.78 &		     -9.41	\\
T40 &	                 M0.5 & 3780 &		     1.2 & 	160 &		    0.55 &		         1.73		     &  0.49 &		     -7.33	\\
T44 &	                 K0.0 & 5110 &		     4.1 & 	160 &		    2.68 &		         2.09		     &  1.65 &		     -6.68	\\
T45 &	                 M0.5 & 3780 &		     3.0 & 	160 &		    0.61 &		         1.82		     &  0.49 &		     -6.95	\\
T45a &	                 K7.0 & 4060 &		     1.1 & 	160 &		    0.34 &		         1.18		     &  0.8	 &	         -9.83	\\
T46 &	                 K7.0 & 4060 &		     1.2 & 	160 &		    0.53 &		         1.47		     &  0.75 &		     -8.7	\\
T48 &	                 M3.0 & 3415 &		     1.2 & 	160 &		    0.16 &		         1.14		     &  0.3	 &	         -7.96	\\
T49 &	                 M3.5 & 3340 &		     1.0 & 	160 &		    0.29 &		         1.61		     &  0.25 &		     -7.41	\\
T5 &	                 M3.0 & 3415 &		     1.4 & 	160 &		    0.53 &		         2.08		     &  0.28 &		     -8.51	\\
T50 &	                 M5.0 & 3125 &		     0.1 & 	160 &		    0.14 &		         1.28		     &  0.17 &		     -9.34	\\
T51B &	                 M2.0 & 3560 &		     0.5 & 	160 &		    0.09 &		         0.789		     &  0.44 &		     -9.07	\\
T52 &	                 K0.0 & 5110 &		     1.0 & 	160 &		    2.55 &		         2.04		     &  1.62 &		     -7.48	\\
TWCha &	                 K7.0 & 4060 &		     0.8 & 	160 &		    0.38 &		         1.25		     &  0.79 &		     -8.86	\\
VWCha &	                 K7.0 & 4060 &		     1.9 & 	160 &		    1.64 &		         2.59		     &  0.67 &		     -7.6	\\
\cutinhead{\hspace{2cm}Orion OB1b}
CVSO104 &	             M2.0 & 3490 &		     0.2 & 	360$\pm$3.9 &	0.37 &		         1.66		     &  0.37 &		     -7.94	\\
CVSO107 &	             M0.5 & 3700 &		     0.3 & 	330$\pm$2.5 &	0.32 &		         1.38		     &  0.53 &		     -7.32	\\
CVSO109 &	             M0.0 & 3767 &		     0.06 & 400$\pm$40.0 &	0.59 &		         1.81		     &  0.5	 &	         -6.76	\\
CVSO146 &	             K6.0 & 4020 &		     0.6 & 	332$\pm$1.7 &	0.8	 &	             1.84		     &  0.86 &		     -8.28	\\
CVSO165A &	             K5.5 & 4221 &		     0.32 & 400$\pm$40.0 &	0.9	 &	             1.78		     &  0.84 &		     -8.15	\\
CVSO165B &	             M1.0 & 3849 &		     1.35 & 400$\pm$40.0 &	0.47 &		         1.55		     &  0.58 &		     -7.78	\\
CVSO176 &	             M3.5 & 3260 &		     1.0 & 	302$\pm$2.9 &	0.34 &		         1.83		     &  0.25 &		     -7.38	\\
CVSO58 &	             K7.0 & 3970 &		     0.8 & 	349$\pm$2.8 &	0.32 &		         1.19		     &  0.81 &		     -7.99	\\
CVSO90 &	             M0.5 & 3700 &		     0.1 & 	338$\pm$3.8 &	0.13 &		         0.88		     &  0.62 &		     -7.909	\\
\cutinhead{\hspace{2cm}Lupus}
2MASSJ16085324-3914401 & M3.0 & 3415$\pm$79.0 &	 1.9 & 	200 &	        0.302$\pm$0.1477  &  1.57$\pm$0.38	 &  0.27$\pm$0.03 &	 -9.79	\\
2MASSJ16100133-3906449 & M6.5 & 2935$\pm$68.0 &	 1.7 & 	200 &	        0.2089$\pm$0.1289 &	 1.77$\pm$0.55	 &  0.14$\pm$0.01 &	 -9.6	\\
EXLup &	                 M0.0 & 3850$\pm$177.0 & 1.1 & 	200	&   	    1.2303$\pm$0.5302 &	 2.49$\pm$0.58	 &  0.48$\pm$0.08 &	 -7.44	\\
Par-Lup3-3 &	         M4.0 & 3270$\pm$75.0 &	 2.2 & 	200	&   	    0.24$\pm$0.11	  &  1.59$\pm$0.37	 &  0.23$\pm$0.02 &	 -9.47	\\
RXJ1556.1-3655 &	     M1.0 & 3705$\pm$171.0 & 1.0 & 	150	&   	    0.2344$\pm$0.1	  &  1.17$\pm$0.27	 &  0.47$\pm$0.08 &	 -7.94	\\
SSTc2d160901.4-392512 &	 M4.0 & 3270$\pm$75.0 &	 0.5 & 	200	&   	    0.148$\pm$0.068	  &  1.25$\pm$0.29	 &  0.23$\pm$0.02 &	 -9.65	\\
SSTc2dJ154508.9-341734 & M5.5 & 3060$\pm$71.0 &	 5.5 & 	150	&   	    0.0575$\pm$0.0283 &	 0.85$\pm$0.21	 &  0.14$\pm$0.01 &	 -8.42	\\
SSTc2dJ160000.6-422158 & M4.5 & 3200$\pm$74.0 &	 0.0 & 	150	&   	    0.0871$\pm$0.0415 &	 0.96$\pm$0.23	 &  0.19$\pm$0.02 &	 -9.81	\\
SSTc2dJ160002.4-422216 & M4.0 & 3270$\pm$75.0 &	 1.4 & 	150	&   	    0.1479$\pm$0.0666 &	 1.2$\pm$0.27	 &  0.22$\pm$0.02 &	 -9.68	\\
SSTc2dJ160026.1-415356 & M5.5 & 3060$\pm$71.0 &	 0.9 & 	150	&   	    0.0661$\pm$0.0397 &	 0.91$\pm$0.27	 &  0.14$\pm$0.01 &	 -9.89	\\
SSTc2dJ160836.2-392302 & K6.0 & 4205$\pm$193.0 & 1.7 & 	200	&   	    1.9499$\pm$0.8633 &	 2.63$\pm$0.63	 &  0.77$\pm$0.1 &	 -7.69	\\
SSTc2dJ160927.0-383628 & M4.5 & 3200$\pm$74.0 &	 2.2 & 	200	&   	    0.1148$\pm$0.0501 &	 1.1$\pm$0.24	 &  0.19$\pm$0.02 &	 -7.95	\\
SSTc2dJ161029.6-392215 & M4.5 & 3200$\pm$74.0 &	 0.9 & 	200	&   	    0.1585$\pm$0.0698 &	 1.29$\pm$0.29	 &  0.2$\pm$0.02 &	 -9.78	\\
SSTc2dJ161243.8-381503 & M1.0 & 3705$\pm$171.0 & 0.8 & 	200	&   	    0.6166$\pm$0.2691 &	 1.91$\pm$0.42	 &  0.41$\pm$0.07 &	 -8.8	\\
SSTc2dJ161344.1-373646 & M5.0 & 3125$\pm$72.0 &	 0.6 & 	200	&   	    0.0692$\pm$0.0305 &	 0.9$\pm$0.2	 &  0.16$\pm$0.02 &	 -8.96	\\
Sz100 &	                 M5.5 & 3057$\pm$70.0 &	 0.0 & 	200	&   	    0.169$\pm$0.078	  &  1.43$\pm$0.33	 &  0.16$\pm$0.01 &	 -9.45	\\
Sz103 &	                 M4.0 & 3270$\pm$75.0 &	 0.7 & 	200	&   	    0.188$\pm$0.087	  &  1.41$\pm$0.3	 &  0.23$\pm$0.02 &	 -9.02	\\
Sz104 &	                 M5.0 & 3125$\pm$72.0 &	 0.0 & 	200	&   	    0.102$\pm$0.047	  &  1.11$\pm$0.26	 &  0.17$\pm$0.01 &	 -9.76	\\
Sz106 &	                 M0.5 & 3777$\pm$174.0 & 1.0 & 	200	&   	    0.098$\pm$0.045	  &  0.72$\pm$0.17	 &  0.62$\pm$0.02 &	 -9.81	\\
Sz108B &	             M5.0 & 3125$\pm$72.0 &	 1.6 & 	200	&   	    0.1514$\pm$0.0813 &	 1.33$\pm$0.36	 &  0.18$\pm$0.02 &	 -9.4	\\
Sz110 &	                 M4.0 & 3270$\pm$75.0 &	 0.0 & 	200	&   	    0.276$\pm$0.127	  &  1.61$\pm$0.37	 &  0.24$\pm$0.02 &	 -8.59	\\
Sz111 &	                 M1.0 & 3705$\pm$171.0 & 0.0 & 	200	&   	    0.33$\pm$0.152	  &  1.4$\pm$0.32	 &  0.61$\pm$0.05 &	 -9.19	\\
Sz112 &	                 M5.0 & 3125$\pm$72.0 &	 0.0 & 	200	&   	    0.191$\pm$0.088	  &  1.52$\pm$0.35	 &  0.19$\pm$0.01 &	 -9.69	\\
Sz113 &	                 M4.5 & 3197$\pm$74.0 &	 1.0 & 	200	&   	    0.064$\pm$0.03	  &  0.83$\pm$0.19	 &  0.19$\pm$0.02 &	 -8.85	\\
Sz114 &	                 M4.8 & 3175$\pm$73.0 &	 0.3 & 	200	&   	    0.312$\pm$0.144	  &  1.82$\pm$0.42	 &  0.21$\pm$0.02 &	 -9.0	\\
Sz115 &	                 M4.5 & 3197$\pm$74.0 &	 0.5 & 	200	&   	    0.175$\pm$0.08	  &  1.36$\pm$0.31	 &  0.20$\pm$0.02 &	 -9.25	\\
Sz117 &	                 M3.5 & 3340$\pm$77.0 &	 0.5 & 	200	&   	    0.4467$\pm$0.1927 &	 2.0$\pm$0.43	 &  0.24$\pm$0.02 &	 -8.64	\\
Sz118 &	                 K5.0 & 4350$\pm$200.0 & 1.9 & 	200	&   	    1.0715$\pm$0.4663 &	 1.82$\pm$0.4	 &  0.94$\pm$0.1 &	 -8.93	\\
Sz123A &	             M1.0 & 3705$\pm$171.0 & 1.25 & 200	&   	    0.203$\pm$0.093	  &  1.1$\pm$0.25	 &  0.65$\pm$0.03 &	 -8.89	\\
Sz123B &	             M2.0 & 3560$\pm$164.0 & 0.0 & 	200	&   	    0.051$\pm$0.024	  &  0.58$\pm$0.13	 &  0.46$\pm$0.04 &	 -9.97	\\
Sz129 &	                 K7.0 & 4060$\pm$187.0 & 0.9 & 	150	&   	    0.3715$\pm$0.16	  &  1.23$\pm$0.27	 &  0.75$\pm$0.08 &	 -8.41	\\
Sz130 &	                 M2.0 & 3560$\pm$164.0 & 0.0 & 	150	&   	    0.16$\pm$0.074	  &  1.03$\pm$0.24	 &  0.44$\pm$0.02 &	 -9.2	\\
Sz131 &	                 M3.0 & 3415$\pm$79.0 &	 1.3 & 	150	&   	    0.1318$\pm$0.0583 &	 1.04$\pm$0.23	 &  0.29$\pm$0.03 &	 -9.26	\\
Sz66 &	                 M3.0 & 3415$\pm$79.0 &	 1.0 & 	150	&   	    0.2$\pm$0.092	  &  1.29$\pm$0.3	 &  0.3$\pm$0.03 &	 -8.6	\\
Sz69 &	                 M4.5 & 3197$\pm$74.0 &	 0.0 & 	150	&   	    0.088$\pm$0.041	  &  0.97$\pm$0.22	 &  0.2$\pm$0.01 &	 -9.5	\\
Sz71 &	                 M1.5 & 3632$\pm$167.0 & 0.5 & 	150	&   	    0.309$\pm$0.142	  &  1.43$\pm$0.33	 &  0.5$\pm$0.06 &	 -9.12	\\
Sz72 &	                 M2.0 & 3560$\pm$164.0 & 0.75 & 150	&   	    0.252$\pm$0.116	  &  1.29$\pm$0.3	 &  0.42$\pm$0.05 &	 -8.69	\\
Sz73 &	                 K7.0 & 4060$\pm$187.0 & 3.5 & 	150	&   	    0.419$\pm$0.193	  &  1.35$\pm$0.31	 &  0.75$\pm$0.08 &	 -8.19	\\
Sz74 &	                 M3.5 & 3342$\pm$77.0 &	 1.5 & 	150	&   	    1.043$\pm$0.48	  &  3.13$\pm$0.72	 &  0.27$\pm$0.02 &	 -7.91	\\
Sz75 &	                 K6.0 & 4205$\pm$193.0 & 0.7 & 	150	&   	    1.4454$\pm$0.626  &	 2.26$\pm$0.53	 &  0.82$\pm$0.1 &	 -7.68	\\
Sz76 &	                 M4.0 & 3270$\pm$75.0 &	 0.2 & 	150	&   	    0.1585$\pm$0.0704 &	 1.24$\pm$0.28	 &  0.23$\pm$0.02 &	 -9.27	\\
Sz77 &	                 K7.0 & 4060$\pm$187.0 & 0.0 & 	150	&   	    0.5495$\pm$0.2428 &	 1.5$\pm$0.36	 &  0.71$\pm$0.09 &	 -8.8	\\
Sz81A &	                 M4.5 & 3200$\pm$74.0 &	 0.0 & 	150	&   	    0.2239$\pm$0.1103 &	 1.54$\pm$0.38	 &  0.2$\pm$0.02 &	 -9.02	\\
Sz81B &	                 M5.5 & 3060$\pm$71.0 &	 0.0 & 	150	&   	    0.1096$\pm$0.0638 &	 1.18$\pm$0.34	 &  0.15$\pm$0.02 &	 -9.68	\\
Sz82 &	                 K5.0 & 4350$\pm$200.0 & 0.9 & 	150	&   	    2.33$\pm$1.0397	  &  2.69$\pm$0.65	 &  XXXX$\pm$0.02 &	 -8.04	\\
Sz83 &	                 K7.0 & 4060$\pm$187.0 & 0.0 & 	150	&   	    1.313$\pm$0.605	  &  2.39$\pm$0.55	 &  1.07$\pm$0.04 &	 -7.21	\\
Sz84 &	                 M5.0 & 3125$\pm$72.0 &	 0.0 & 	150	&   	    0.122$\pm$0.056	  &  1.21$\pm$0.28	 &  0.17$\pm$0.01 &	 -9.24	\\
Sz88A &	                 M0.0 & 3850$\pm$177.0 & 0.25 & 200	&   	    0.488$\pm$0.225	  &  1.61$\pm$0.37	 &  0.74$\pm$0.05 &	 -8.2	\\
Sz88B &	                 M4.5 & 3197$\pm$74.0 &	 0.0 & 	200	&   	    0.118$\pm$0.054	  &  1.12$\pm$0.26	 &  0.2$\pm$0.01 &	 -9.74	\\
Sz90 &	                 K7.0 & 4060$\pm$187.0 & 1.8 & 	200	&   	    0.6607$\pm$0.2845 &	 1.64$\pm$0.36	 &  0.69$\pm$0.09 &	 -8.66	\\
Sz91 &	                 M1.0 & 3705$\pm$171.0 & 1.2 & 	200	&   	    0.311$\pm$0.143	  &  1.36$\pm$0.31	 &  0.6$\pm$0.05 &	 -8.78	\\
Sz95 &	                 M3.0 & 3415$\pm$79.0 &	 0.8 & 	200	&   	    0.4169$\pm$0.1842 &	 1.84$\pm$0.41	 &  0.28$\pm$0.03 &	 -9.13	\\
Sz96 &	                 M1.0 & 3705$\pm$171.0 & 0.8 & 	200	&   	    0.6918$\pm$0.3234 &	 2.02$\pm$0.47	 &  0.4$\pm$0.07 &	 -9.07	\\
Sz97 &	                 M4.0 & 3270$\pm$75.0 &  0.0 & 	200	&   	    0.169$\pm$0.078	  &  1.34$\pm$0.28	 &  0.23$\pm$0.02 &	 -9.54	\\
Sz98 &	                 K7.0 & 4060$\pm$187.0 & 1.0 & 	200	&   	    2.5119$\pm$1.0755 &	 3.2$\pm$0.69	 &  0.65$\pm$0.07 &	 -7.24	\\
Sz99 &	                 M4.0 & 3270$\pm$75.0 &	 0.0 & 	200	&   	    0.074$\pm$0.034	&    0.89$\pm$0.2	 &  0.21$\pm$0.02 &	 -9.37	\\
\cutinhead{\hspace{2cm}Upper Scorpius}
J15354856-2958551 E &    M4.5 & 3200 &		     0.0 & 	145$\pm$0.0 &	0.1  &               1.03		     &  0.2	 &	         -9.45	\\
J15354856-2958551 W &    M4.5 & 3200 &		     0.0 & 	145$\pm$0.0 &	0.1  &               1.03		     &  0.2	 &	         -9.56	\\
J15530132-2114135 &      M4.5 & 3200 &		     0.8 & 	146$\pm$2.0 &	0.05 &               0.73		     &  0.19 &		     -9.82	\\
J15534211-2049282 &      M4.0 & 3270 &		     1.2 & 	136$\pm$4.0 &	0.09 &               0.93		     &  0.24 &		     -9.44	\\
J15582981-2310077 &      M4.5 & 3200 &		     1.0 & 	147$\pm$3.0 &	0.05 &               0.73		     &  0.19 &		     -9.15	\\
J16001844-2230114 &      M4.5 & 3200 &		     0.8 & 	138$\pm$9.0 &	0.08 &               0.92		     &  0.2	 &	         -8.69	\\
J16014086-2258103 &      M3.0 & 3415 &		     1.2 & 	145$\pm$0.0 &	0.12 &               0.99		     &  0.31 &		     -8.13	\\
J16024152-2138245 &      M5.5 & 3060 &		     0.6 & 	142$\pm$2.0 &	0.03 &               0.62		     &  0.12 &		     -9.56	\\
J16035767-2031055 &      K6.0 & 4205 &		     0.7 & 	143$\pm$1.0 &	0.48 &               1.31		     &  0.91 &		     -9.06	\\
J16041893-2430392 &      M2.0 & 3560 &		     0.3 & 	145$\pm$0.0 &	0.45 &               1.76		     &  0.37 &		     -9.83	\\
J16054540-2023088 &      M4.5 & 3200 &		     0.6 & 	145$\pm$2.0 &	0.1	&	             1.03		     &  0.2	 &	         -9.45	\\
J16062196-1928445 &      M1.0 & 3705 &		     0.8 & 	145$\pm$0.0 &	0.34 &		         1.42		     &  0.46 &	         -8.22	\\
J16064385-1908056 &      K7.0 & 4060 &		     0.4 & 	144$\pm$7.0 &	0.29 &		         1.09		     &  0.82 &	         -9.58	\\
J16072625-2432079 &      M3.0 & 3415 &		     0.7 & 	143$\pm$2.0 &	0.18 &		         1.21		     &  0.29 &	         -9.34	\\
J16082324-1930009 &      M0.0 & 3850 &		     1.1 & 	138$\pm$1.0 &	0.32 &		         1.27		     &  0.61 &	         -9.10	\\
J16082751-1949047 &      M5.5 & 3060 &		     0.6 & 	145$\pm$0.0 &	0.06 &		         0.87		     &  0.14 &	         -9.71	\\
J16090075-1908526 &      M0.0 & 3850 &		     1.0 & 	138$\pm$1.0 &	0.32 &		         1.27		     &  0.6	 &	         -8.76	\\
J16111330-2019029 &      M3.5 & 3340 &		     0.6 & 	155$\pm$1.0 &	0.03 &		         0.52		     &  0.27 &		     -9.01	\\
J16123916-1859284 &      M1.0 & 3705 &		     0.6 & 	139$\pm$2.0 &	0.22 &		         1.14		     &  0.5	 &	         -9.32	\\
J16133650-2503473 &      M3.0 & 3415 &		     1.0 & 	145$\pm$0.0 &	0.11 &		         0.95		     &  0.32 &		     -8.53	\\
J16135434-2320342 &      M4.5 & 3200 &		     0.3 & 	145$\pm$0.0 &	0.12 &		         1.13		     &  0.2	 &	         -8.93	\\
J16143367-1900133 &      M3.0 & 3415 &		     1.9 & 	142$\pm$2.0 &	0.52 &		         2.06		     &  0.29 &		     -9.29	\\
J16154416-1921171 &      K7.0 & 4060 &		     2.8 & 	132$\pm$2.0 &	0.3 &		         1.11		     &  0.81 &		     -7.61	\\
\enddata
\tablecomments{This table lists the stellar parameters for the Classical T Tauri Stars (CTTS) sample. Columns include: object name, spectral type ($SpT$), effective temperature ($T_{\rm eff}$), stellar luminosity ($L$), radius ($R$), mass ($M$), Gaia-based distance ($d$) with error ($\sigma_{\rm err}$), visual extinction ($A_V$), and logarithmic mass accretion rate (log($\dot{M}$)). These parameters are as reported in \citet{alcala2014,alcala2017} (Lup), \citet{manara2016, manara2017b} (Cha I),  \citet{pittman2022} for \mdot\ and \citet{manara2021} for the rest (OB1b), and \citet{manara2020} (USco)}

\end{deluxetable*}

%% file: table_wtts.tex
\begin{deluxetable*}{ccccccc}[h]
    \tablecaption{Stellar parameters of the WTTS sample}\label{table:wtts_table}
    \tablewidth{0pt}
    \tablehead{
    \colhead{Object} & 
    \colhead{SpT} &
    \colhead{$d$ [pc]} &
    \colhead{M [$M_\odot$]} &
    \colhead{$T_{eff}$ [K]} &
    \colhead{Log $L/L_{\odot}$} &
    \colhead{Ref}
    }
\startdata
    TWA9A          & K5   & 68           & 0.81          & 4350              & -0.61             & M13  \\
    RXJ1540.7-3756 & K6   & 150          & --            & 4205              & -0.41             & M17a \\
    SO879          & K7   & 360          & 1.07          & 4060              & -0.29             & M13  \\
    TWA25          & M0   & 54           & 0.84          & 3850              & -0.61             & M13  \\
    TWA14          & M0.5 & 96           & 0.73          & 3780              & -0.83             & M13  \\
    TWA13B         & M1   & 59           & 0.68          & 3705              & -0.70             & M13  \\
    TWA2A          & M2   & 47           & 0.55          & 3560              & -0.48             & M13  \\
    TWA9B          & M3   & 68           & 0.37          & 3415              & -1.17             & M13  \\
    TWA15A         & M3.5 & 111          & 0.30          & 3340              & -0.95             & M13  \\
    Sz94           & M4   & 200          & 0.28          & 3270              & -0.76             & M13  \\
    SO797          & M4.5 & 360          & 0.19          & 3200              & -1.26             & M13  \\
    Par-Lup3-2     & M5   & 200          & 0.18          & 3125              & -0.75             & M13  \\
    SO999          & M5.5 & 360          & 0.13          & 3060              & -1.28             & M13  \\
\enddata
    \tablecomments{This table lists the stellar parameters of Weak T Tauri Stars (WTTS) in the sample. Parameters include spectral type (SpT), distance ($d$), mass (M), effective temperature ($T_{eff}$), and luminosity (Log $L/L_{\odot}$). Stars with data obtained from \citet{manara2013b} and \citet{manara2017a} are identified with M13 and M17a, respectively.}
    
\end{deluxetable*}